\documentclass[journal,twocolumn]{IEEEtran}
\usepackage{epsfig,makeidx,color}
\usepackage{amsmath,amssymb,amsthm,bbm}
\usepackage{cite,graphicx}
\usepackage{enumerate}

\def\cL{{\cal L}}

\def\cH{{\cal H}}

\def\cX{{\cal X}}

\def\cQ{{\cal Q}}

\def\rH{{\rm H}}

\def\rT{{\rm T}}

\def\uE{{\mathbb E}}

\DeclareMathOperator*{\argmin}{\arg\!\min}
\DeclareMathOperator*{\argmax}{\arg\!\max}

\newtheorem{mylemma}{\bf Lemma} 

\def\be{ \begin{equation} }
\def\ee{ \end{equation} }
\def\bea{ \begin{eqnarray} }
\def\eea{ \end{eqnarray} }
\def\bx{{\bf x}}
\def\by{{\bf y}}

\def\bq{{\bf q}}

\def\bs{{\bf s}}
\def\ba{{\bf a}}

\def\bn{{\bf n}}
\def\bh{{\bf h}}

\def\bee{{\bf e}}

\def\bA{{\bf A}}
\def\bB{{\bf B}}

\def\bD{{\bf D}}

\def\bH{{\bf H}}
\def\bI{{\bf I}}

\def\bV{{\bf V}}

\def\bR{{\bf R}}
\def\bW{{\bf W}}

\def\b0{{\bf 0}}

\def\bDelta{{\bf \Delta}}

\def\cC{{\cal C}}

\def\cI{{\cal I}}
\def\cN{{\cal N}}
\def\cS{{\cal S}}

\ifCLASSOPTIONonecolumn
  \newcommand{\figwidth}{0.50\columnwidth}
\else
  \newcommand{\figwidth}{0.95\columnwidth}
\fi

\begin{document}

\title{A Variational Inference based Detection Method 
for Repetition Coded Generalized Spatial Modulation}

\author{Jinho Choi\\
\thanks{The author is with
the School of Information Technology,
Deakin University,
Burwood, VIC 3125, Australia
(e-mail: jinho.choi@deakin.edu.au).}}

\date{today}
\maketitle

\begin{abstract}
In this paper, we consider a simple coding 
scheme for spatial modulation (SM), where the same set of
active transmit antennas is repeatedly used 
over consecutive multiple transmissions.
Based on a Gaussian approximation, an approximate
maximum likelihood (ML) detection problem is formulated 
to detect the indices of active transmit antennas.
We show that the solution to the approximate ML detection problem
can achieve a full coding gain.
Furthermore, we develop a low-complexity iterative algorithm
to solve the problem with low complexity 
based on a well-known machine learning approach,
i.e., variational inference. Simulation results show that
the proposed algorithm can have a near ML performance.
A salient feature
of the proposed algorithm
is that its complexity is independent of 
the number of active transmit antennas, 
whereas an exhaustive search for the ML 
problem requires a complexity that grows exponentially
with the number of active transmit antennas. 
\end{abstract}

{\IEEEkeywords
Spatial Modulation;
Index Modulation;
Repeated Transmit Diversity;
Variational Inference}

\ifCLASSOPTIONonecolumn
\baselineskip 26pt
\fi

\section{Introduction} \label{S:Intro}

In wireless communications,
multiple-input multiple-output (MIMO) systems play a crucial
role in improving the spectral efficiency by exploiting
multiple antennas \cite{Telatar, Foschini98} and have been 
actively considered for cellular systems, 
e.g., long-term evolution (LTE)
\cite{DahlmanBook11} and fifth generation (5G) systems
\cite{Boccardi14}.

There have been various transmission schemes over MIMO channels
that have different advantages and characteristics
due to the fundamental tradeoff between diversity
and multiplexing gain of MIMO channels \cite{Zheng03}.
In \cite{Mesleh08}, a simple but efficient approach to exploit
the spatial multiplexing gain of MIMO channels
is proposed without channel state
information (CSI) at the transmitter.
This approach, which is called spatial modulation (SM),
considers multiple transmit antennas as
additional constellation points and uses one transmit antenna
at a time, 
which requires a single radio frequency (RF) chain.
As a result, although there can be a number of transmit
antennas, the transmitter' cost can be low.
In \cite{Jeganathan08}, the optimal detection
for SM is studied with performance analysis.
In \cite{Rajashekar14} \cite{Men14}, 
it is shown that the complexity of the maximum likelihood (ML)
detection involving joint detection
of the transmit antenna index and of the transmitted
symbol can be low if a square-quadrature amplitude
modulation (QAM) or phase shift keying (PSK) is employed.
With multiple RF chains, SM can be 
further generalized where more than one transmit antenna
can be active at a time as
an index modulation (IM) scheme in the space
domain \cite{Younis10, Renzo13, Renzo14, Basar16}. 
The resulting SM is called generalized SM. Throughout the paper,
we consider generalized SM and call it SM for convenience
(i.e., SM implies GSM in this paper).

Since the detection performance of active transmit antennas
in SM is limited, various
transmit diversity schemes or channel
coding approaches are considered. 
For example, in \cite{Mesleh10, Basar11, Choi14}, 
various designs for trellis coded SM  (TCSM) are considered 
with channel coding for SM, 
and in \cite{Basar11b, Li14}, 
space-time block coding is applied to SM.
Spatial and temporal modulation schemes based on 
space-time block coding are studied 
for a high diversity gain and a low error rate
in \cite{Rajashekar12, Helmy16}.
Since SM is an IM scheme,
certain transmit diversity techniques used for other IM
schemes such as 
orthogonal frequency division multiplexing (OFDM) with IM
\cite{Basar13} can be applied.
Among them, we can consider a simple 
but effective approach studied in \cite{Choi17},
where the same set of active indices is repeatedly used 
over consecutive multiple transmissions in order to provide a coding
gain in detecting active indices.

In this paper, we apply the transmit scheme 
considered in \cite{Choi17} to SM. 
Throughout the paper, under the assumption of perfect CSI at a receiver,
in order to detect the indices of active transmit antennas
with low computational complexity, we propose an iterative
algorithm based on a well-known machine learning approach,
i.e., variational inference \cite{Bishop06, Blei17}. 
The proposed iterative algorithm
is to find an approximate solution to
an ML detection problem
that is formulated using a Gaussian approximation.
Thanks to the Gaussian approximation,
it is possible to find low-complexity updating rules
for the proposed iterative algorithm
and the resulting computational complexity
becomes independent of the number of active transmit antennas.
This feature is attractive since the optimal ML approach
has a computational complexity that grows exponentially with 
the number of active transmit antennas.

It is noteworthy that an ML approach 
is also studied in \cite{Lin15, Freud18}. 
In particular, as in \cite{Lin15},
we focus on the detection of active transmit antennas assuming
that the data symbols of active antennas are Gaussian
(which is a Gaussian approximation).
However, in this paper,
the activity of each transmit antenna is 
modeled by an independent Bernoulli random variable to avoid
an exhaustive search using
the variational inference for the detection of active antennas.
On the other hand, an exhaustive
search for the detection of active antennas
is considered in \cite{Lin15, Freud18}. 

The main contributions of the paper are two-fold. First,
we propose a transmit coding scheme for SM
to allow a reliable detection of IM bits and show that
a full transmit coding gain can be achieved
by deriving an approximate expression for the probability of index
error. Secondly, in order to lower the computational complexity
of the detection, we consider a two-step approach and propose a 
low-complexity iterative algorithm for the first step based on
variational inference. 
The resulting approach can also 
provide an approximate solution to the ML problem
in \cite{Lin15, Freud18} with the complexity that
grows linearly with the number of transmit antennas and is independent
of the number of active transmit antennas.
On the other hand, the complexity of the approaches
in \cite{Lin15, Freud18} grows exponentially with 
the number of active transmit antennas.

The rest of the paper is organized as follows.
Section~\ref{S:SM} presents a system model for 
SM with a transmit coding scheme. Based on 
a Gaussian approximation, we formulate an approximate ML problem
to detect the indices of active transmit antennas in Section~\ref{S:ML}.
In Section~\ref{S:CAVI}, we apply
variational inference to the approximate ML problem and develop
an iterative algorithm.
A performance analysis is carried out in Section~\ref{S:PA}
to find the probability of index error of 
the approximate ML problem. We 
present simulation results in Section~\ref{S:Sim}
and conclude the paper with some remarks in Section~\ref{S:Conc}.

{\it Notation}:
Matrices and vectors are denoted by upper- and lower-case
boldface letters, respectively.
The superscripts $\rT$ and $\rH$
denote the transpose and complex conjugate, respectively.
The $p$-norm of a vector $\ba$ is denoted by $|| \ba ||_p$
(If $p = 2$, the norm is denoted by $||\ba||$ without
the subscript).
The support of a vector is denoted by ${\rm supp} (\bx)$
(which is the index set of the non-zero elements of $\bx$).
$\uE[\cdot]$
and ${\rm Var}(\cdot)$
denote the statistical expectation and variance, respectively.
$\cC \cN(\ba, \bR)$
represents the distribution of
circularly symmetric complex Gaussian (CSCG)
random vectors with mean vector $\ba$ and
covariance matrix $\bR$.

\section{System Model} \label{S:SM}

Suppose that there are $L$ transmit and $N$ receive antennas
and denote by $\bH$ the MIMO channel matrix of size $N \times L$.
Let $s_l$ denote the signal to be transmitted
through the $l$th transmit antenna.
The received signal vector is given by
\be
\by = \bH \bs + \bn,
\ee
where 
$\bs = [s_1 \ \ldots \ s_L]^\rT$ is the symbol vector
and $\bn \sim \cC \cN(0, N_0\bI)$ is the background noise.
In SM \cite{Renzo14}, there are a few RF chains, say $K$, and only
$K$ symbols are non-zero, while the other symbols are zero.
For convenience, 
the transmit
antennas corresponding to non-zero symbols are referred to
as active transmit antennas
and $K$ is referred to as
the number of active transmit antennas.
Clearly, $K \le L$. In addition, let 
$\cI$ denote the set of the indices of active transmit antennas,
while the indices of active transmit antennas 
are referred to as active indices for convenience.
In addition, a non-zero symbol, $|s_l| > 0$, is called 
data modulation (DM) symbol.
Denote by $\cS$ the alphabet of DM symbols. 
The mean and variance of a non-zero DM symbol, i.e., $|s_l| > 0$,
are assumed to be 0 and $\sigma_s^2$, respectively.
For equally likely DM symbols, we have
$\sigma_s^2 = \uE[|s|^2] = \frac{1}{|\cS|} \sum_{s \in \cS} |s|^2$.
While DM symbols
can represent information bits, which are referred to as 
DM bits,
as in conventional modulation
schemes, the index set of non-zero symbols or 
the indices of active transmit antennas 
can also represent information bits that are referred to IM bits.
The number of IM bits is given by
$B_{\rm im} = \lfloor \log_2 \binom{L}{K} \rfloor$.

In general, provided that the set of active indices 
is correctly decided,
the detection of DM symbols can be reliably performed using
various MIMO detectors
for a sufficiently small $K$ \cite{VerduBook,ChoiJBook2}.
On the other hand, the 
detection performance of active indices or IM bits
can be relatively poor (especially, when $L > N$) for low
and moderate 
signal-to-noise ratio (SNR)
\cite{Choi14}.
To avoid this problem, 
we can consider a modification of SM with multiple symbol
vectors based on the approach in \cite{Choi17}.
Suppose that a slot consists of $M$ consecutive symbol vectors,
where $M$ is referred to as the slot length.
For each symbol vector in a slot, different DM bits are transmitted,
while the IM bits are fixed.
In this case, the $m$th received signal vector can be given by
\be
\by_m = \bH \bs_m + \bn_m, \ m = 1,\ldots, M,
	\label{EQ:ym}
\ee
where $\bs_m$ is the $m$th symbol vector
and $\bn_m \sim \cC \cN(0, N_0\bI)$ is the background noise.
Since the same IM bits are transmitted, we have
\be
\cI = 
{\rm supp} (\bs_m), \ m =1, \ldots, M.
\ee
The resulting SM 
with repeated uses of the same set of active indices
is referred to as repetition coding
SM (RCSM) in this paper.
The total number of bits per slot that can be transmitted by RCSM
becomes
$B_{\rm total} = M K \log_2 |\cS| + \lfloor \log_2 \binom{L}{K} \rfloor$,
where $|\cS|$ is assumed to be a power of 2, and 
the number of bits per symbol becomes 
\be
B_{\rm sym} = K \log_2 |\cS| +
\frac{ \lfloor \log_2 \binom{L}{K} \rfloor}{M}.
	\label{EQ:Bsym}
\ee
Clearly, the spectral efficiency decreases with $M$.
However, we expect to have a higher coding gain 
for a large $M$, which results in a more reliable detection
for IM bits.

\section{ML Approach to Index Detection}	 \label{S:ML}

In this section, we consider an ML detection approach
for IM bits or active indices in RCSM.
Throughout the paper, we assume that the receiver has perfect CSI.

Prior to the discussion of 
an ML approach to detect the set of 
active indices,
we consider a simple low-complexity detection method 
based on the correlators.
The detection of active indices
is similar to the detection of the spreading codes that
are used by multiple users in the context of
code division multiple access (CDMA).
Assuming that each column of $\bH$ as a spreading code,
the energy of the correlator can be used to decide whether or
not the corresponding transmit antenna is active.
To this end,
the energy of the $l$th correlator can be defined as
$\rho_l = \sum_{m=1}^M |\bh_l^\rH \by_m|^2$,
where $\bh_l$ denotes the $l$th column vector of $\bH$.
Denote by $\hat l(k)$ the index 
corresponding to the $k$th largest energy of the correlator
output. Then,
the set of active indices can be estimated as
\be
\hat \cI_{\rm corr} = \{ \hat l(1), \ldots, \hat l(K)\}.
\ee
This approach could provide a reasonably good performance
if the $\bh_l$'s are orthogonal or $N \gg K$.
The computational complexity of 
the correlator based detector
is $O(LN)$. 

If $N$ is not sufficiently large, 
the correlator based detector cannot provide a good performance.
Thus, we may need to consider an optimal approach
\cite{Jeganathan08, Renzo14}.  
For convenience, let 
$$
x_l = \left\{
\begin{array}{ll}
1, & \mbox{if $|s_{m,l}| > 0$} \cr
0, & \mbox{if $|s_{m,l}| = 0$,} \cr
\end{array}
\right.
$$
where $s_{m,l}$ represents the $l$th element of $\bs_m$.
Let $\bx = [x_1 \ \ldots \ x_L]^\rT$, which is referred to
as the index vector. In addition, define
\be
\cX_K = \{\bx \,\bigl|\, ||\bx||_0 = K, \ x_l \in \{0,1\}\}.
\ee
The ML detection problem can be formulated as
\begin{align}
\{\hat \bs_m, \bx\} 
& = \argmax_{\bx \in \cX_K,[\bs_m]_\bx \in \cS^K} 
f(\by\,|\,\{ \bs_m\}, \bx)  \cr
& = \argmax_{\bx \in \cX_K} \max_{[\bs_m]_\bx \in \cS^K} 
f(\by\,|\,\{ \bs_m\}, \bx) ,
	\label{EQ:oml}
\end{align}
where 
$f(\by\,|\,\{ \bs_m\}, \bx)$
is the likelihood function of $\{ \bs_m\}$ and $\bx$
for given $\by = [\by_1^\rT \ \ldots \ \by_M^\rT]^\rT$ and
$[\bs]_\bx$ 
represents the subvector of $\bs$ obtained by taking the elements 
corresponding to the non-zero elements of $\bx$.
Unfortunately, the complexity to solve \eqref{EQ:oml}
is prohibitively high.
To lower the complexity,
we may consider a suboptimal two-step approach.
In the first step, we detect 
the support of $\bs_m$, i.e., $\cI$.
Once an estimate of $\cI$ is available, 
the detection of the non-zero elements of $\bs_m$ are carried
out in the second step.
Since the second step is the conventional MIMO detection,
throughout the paper, we mainly focus on the first step,
which is referred to as the support or index detection.

For the first step,
we can consider the ML detection of
$\cI$ or $\bx$ for given $\by$ as follows:
\be
\hat \bx = \argmax_{\bx \in \cX_K}
f(\by\,|\, \bx) ,
\ee
where
\begin{align}
f(\by\,|\, \bx) = \prod_{m =1}^M \sum_{\bs_m}
f(\by_m\,|\, \bs_m, \bx) \Pr(\bs_m\,|\, \bx) .
	\label{EQ:lhx}
\end{align}
The likelihood function of $\bx$ in \eqref{EQ:lhx}
is a Gaussian mixture. Thus,
the computational
complexity to find the likelihood function of $\bx$ for given $\by$
in  \eqref{EQ:lhx} can be high due to the summation over
$\bs_m$. To avoid this problem, we resort to an approximation.

For given $\bx$,
$\bs_m$ can also be expressed as
\be
\bs_m = \bD_m \bx,
\ee
where $\bD_m = {\rm diag}(d_{m,1}, \ldots, d_{m,L})$ is a diagonal matrix.
Clearly, $d_{m,l} = s_{m,l}$ if $x_l = 1$. Otherwise, 
$d_{m,l}$ can be any non-zero element in $\cS$.
Let us employ an Gaussian approximation 
where the $d_{m,l}$'s are assumed to be CSCG random variables
with zero-mean and variance $\sigma_s^2$.
Furthermore, for convenience, we assume that $\sigma_s^2 = 1$,
while $N_0$ becomes $\gamma^{-1}$, where 
$\gamma$ represents the SNR, which is given by
$\gamma = \frac{\uE[|s|^2]}{N_0}$.
Based on the Gaussian approximation, for given $\bx$ and $\bH$, we have
\begin{align}
\by_m \,|\, \bx \sim 
f(\by_m\,|\, \bx) & \approx  f_{\rm ga} (\by_m\,|\, \bx) \cr
& = \cC \cN(\b0,  \bH \bR_\bx \bH^\rH +
\gamma^{-1} \bI),
	\label{EQ:GA}
\end{align}
where 
\be
\bR_\bx = {\rm diag}(x_1, \ldots, x_L).
\ee
Then, the index detection
based on the approximate likelihood function in \eqref{EQ:GA}
can be carried out as follows:
\begin{align}
\hat \bx_{\rm ga}
& = \argmax_{\bx \in \cX_K} \prod_{m=1}^M f_{\rm ga} (\by_m \,|\, \bx) \cr
& = \argmin_{\bx \in \cX_K} \sum_{m=1}^M 
\by_m^\rH \bV (\bx) \by_m + M \phi(\bx),
	\label{EQ:ml_ga}
\end{align}
where 
$\bV(\bx) = \left(\bH \bR_\bx \bH^\rH + 
\gamma^{-1} \bI \right)^{-1}$ and
$\phi(\bx) = \ln\det( \bH \bR_\bx \bH^\rH + \gamma^{-1} \bI )$.
The detection problem in \eqref{EQ:ml_ga}
is also considered (with $M = 1$) in \cite{Lin15, Freud18}.

In RCSM, we expect to have a coding gain of $M$ in detecting 
the set of active indices, $\cI$, or $\bx$. 
However, in \eqref{EQ:ml_ga},
since the approximate ML formulation is considered to lower the complexity,
it is not clear whether or not 
the detection based on \eqref{EQ:ml_ga} can exploit a full coding gain.
Fortunately, as will be shown in Section~\ref{S:PA},
the solution to \eqref{EQ:ml_ga} 
can achieve a full transmit coding gain, $M$.

Thanks to the Gaussian approximation,
the complexity to find the likelihood for given $\bx$
can be low compared to that in \eqref{EQ:lhx}.
However, the approximate ML formulation in \eqref{EQ:ml_ga}
requires a complexity that is proportional to $|\cX_K|$
if an exhaustive search is used.
In the next section, we apply a well-known machine learning approach
to find an approximate solution to 
\eqref{EQ:ml_ga} with low complexity.

\section{Application of Variational Inference for Index Detection}
\label{S:CAVI}

In this section, we consider a variational inference approach
\cite{Bishop06, Blei17} 
for the index detection to solve \eqref{EQ:ml_ga} with low complexity,
where a soft-decision of $\bx$ 
(rather than a hard-decision) is considered.

\subsection{Variational Inference based Detection}

We assume that the elements of $\bx$ 
are independent random variables in \eqref{EQ:ml_ga}, 
which are seen as the latent variables.
In particular, we consider the following assumption.
\begin{itemize}
\item[{\bf A0)}] The $x_l$'s are independent
binary random variables and denote by
$q_l$ the probability that $x_l = 1$, i.e.,
$q_l = \Pr(x_l =1)$,
which is referred to as the variational probability.
\end{itemize}

In the context of variational inference \cite{Jordan99, Bishop06, Blei17},
the estimation of $\bx$ becomes the following optimization
problem:
\be
\bq^* = \argmin_{ \bq \in \cQ}
{\rm KL} \left(\bq || \Pr( \bx\,|\,\by) \right),
	\label{EQ:KL}
\ee
where $\cQ$ represents the collection
of all the possible distributions of $\bq = [q_1 \ \ldots \ q_L]^\rT$,
$\Pr( \bx \,|\,\by)$ is the a posteriori probability
of $\bx$, and ${\rm KL}(\cdot)$ 
is the Kullback-Leibler (KL) divergence \cite{CoverBook}.
Then, from $\bq^*$,
the estimate of $\bx$ can be found as
\be
\hat x_l = 
\left\{
\begin{array}{ll}
1, \ \mbox{if $q_l^*$ is one of the $K$ largest values of $\bq^*$} \cr
0, \ \mbox{o.w.} \cr
\end{array}
\right.
\ee

As shown in \cite{Blei17},
the minimization of the KL divergence in \eqref{EQ:KL}
is equivalent to the maximization of the evidence lower bound (ELBO),
which is given by
\be
{\rm ELBO} (\bq) = \uE[\ln f(\by, \bx)] - \uE[\ln \Pr(\bx)],
\ee
where the expectation is carried out over $\bx$
under the assumption of {\bf A0}.
Let $\bx_{-l} = [x_1 \ \ldots \ x_{l-1} \ x_{l+1} \ \ldots \ x_L]^\rT$.
Then, for given $\bx_{-l}$,
it can be shown that
\be
q_l \propto 
\exp \left( \uE_{-l} [\ln f(x_l\,|\, \bx_{-l}, \by) ] \right),
	\label{EQ:qE}
\ee
where the expectation,
denoted by $\uE_{-l}$, is carried out over $\bx_{-l}$.
The coordinate ascent variational inference (CAVI) algorithm
\cite{Bishop06, Blei17} is to update $q_l$, $l = 1,\ldots, L$,
with fixing the other variational distributions, $\bq_{-l}$,
through iterations.

To perform the CAVI algorithm,
we need to find 
$\Pr(x_l\,|\, \by, \bx_{-l})$ or
$\ln \Pr(x_l\,|\, \by, \bx_{-l})$.
It can be shown that
\begin{align}
\ln \Pr(x_l\,|\, \by, \bx_{-l})
& = \ln \frac{f(\bx, \by)}{f(\by, \bx_{-l})} 
= \ln \frac{f(\bx, \by)}{\sum_{x_l} f(\by, \bx)} \cr
& = \ln f(\bx, \by)+ {\rm const.} \cr
& = \ln f(\by|\bx_{-l}, x_l) + {\rm const.},
\end{align}
where the third equality is due to the fact that
${\sum_{x_l} f(\by, \bx)}$ does not depend on $x_l$
(and can be treated as a constant) and
the fourth equality is due to the assumption that
$\Pr(\bx)$ is the same for all $\bx \in \cX_K$.
Thus, for the updating in \eqref{EQ:qE},
we can carry out the expectation over $\bx_{-l}$ as follows:
\begin{eqnarray}
\uE_{-l} [\ln \Pr(x_l\,|\, \by, \bx_{-l})]
& = & \uE_{-l} [\ln f(\by\,|\, \bx_{-l}, x_l)] + C_1 \cr
& \approx & - \sum_{m=1}^M \by_m^\rH \uE_{-l} [\bV(\bx)] \by_m \cr
& & - M \uE_{-l} [\phi(\bx)] + C_2,
	\label{EQ:EE}
\end{eqnarray}
where the approximation is due to the Gaussian approximation 
as in \eqref{EQ:ml_ga},
and $C_1$ and $C_2$ are constants.

As shown in \eqref{EQ:EE},
we need to obtain $\uE_{-l} [\bV(\bx)]$
and $\uE_{-l} [\phi(\bx)]$
as closed-form expressions to perform \eqref{EQ:qE}.
Unfortunately, it is not easy to find
closed-form expressions. However, we can find tight
bounds in the following lemma.

\begin{mylemma}	\label{L:Ineq2}
Under the assumption of {\bf A0},
we have
\begin{align}
\uE_{-l} [\bV(\bx)] 
& \succeq \bR_l (x_l)^{-1} \cr
\uE_{-l} [\phi(\bx)] 
& \le \ln \det  \bR_l (x_l),
	\label{EQ:L1}
\end{align}
where 
$\bR_l (x_l) = 
\bh_l \bh_l^\rH x_l + \sum_{t \ne l}
\bh_t \bh_t^\rH q_t + \gamma^{-1} \bI$ and $\bA \succeq \bB$
implies that $\bA - \bB$ is 
positive semidefinite.
The inequalities in \eqref{EQ:L1} are tight 
if $q_t \to 0$ or $1$ for $t \in \{1, \ldots, l-1,l+1, \ldots, L\}$.
\end{mylemma}
\begin{IEEEproof}
See Appendix~\ref{A:Ineq2}.
\end{IEEEproof}

For the CAVI algorithm with the approximations in \eqref{EQ:L1},
let 
\begin{align}
\bR_l^{(i)} (x_l)
& = 
\bh_l \bh_l^\rH x_l + \sum_{t <l } \bh_t \bh_t^\rH q_t^{(i)} \cr
& \quad 
+ \sum_{t >l } \bh_t \bh_t^\rH q_t^{(i-1)} 
+ \gamma^{-1} \bI ,
	\label{EQ:iRl}
\end{align}
where the superscript $(i)$ represents the $i$th iteration
(i.e., $i$ is used for the iteration index).
Then, the term on the right-hand side (RHS) in \eqref{EQ:qE}
can be approximated by
\begin{align}
\chi_l^{(i)} (x_l) & = e^{
- \sum_{m=1}^M \by_m^\rH \bR_{l}^{(i)} (x_l)^{-1} \by_m 
- M \ln \det(\bR_l^{(i)} (x_l))},  \cr
& \qquad x_l \in \{0,1\},
	\label{EQ:chil}
\end{align}
which is an estimate of 
$e^{ \uE_{-l} [\ln f(x_l\,|\, \bx_{-l}, \by) ]}$
in \eqref{EQ:qE} with the approximations in \eqref{EQ:L1}.
For the normalization,
let
$\bar \chi_l^{(i)} = 
\frac{\chi_l^{(i)}(1)}{\chi_l^{(i)}(0) + \chi_l^{(i)}(1)}$.
Then, the updating rule for the CAVI algorithm can be given by
\be
q_l^{(i)} = (1-\mu) q_l^{(i-1)} + \mu \bar \chi_l^{(i)},
\ l = 1, \ldots, L,
\ee
where $\mu \in [0,1]$ is the step-size. 
In each iteration,
the updating is carried out
in the ascending order, i.e., from $l = 1$ to $L$.

After a number of iterations, we can decide
the set of active indices with 
the $x_l$'s corresponding to the $K$
largest values of $q_l^{(N_{\rm run})}$,
where $N_{\rm run}$ denotes the number of iterations of the
CAVI algorithm. The corresponding set
of estimated active indices
is denoted by $\hat \cI_{\rm cavi}$.

\subsection{Low-Complexity Updating Rules}	\label{SS:LC}

From \eqref{EQ:chil},
we can see that the computational complexity
of the CAVI algorithm
depends on the complexity to perform the matrix
inversion (i.e, $\bR_{l}^{(i)} (x_l)^{-1}$)
and the determinant
(i.e, $\det (\bR_{l}^{(i)} (x_l))$).
Since the size of $\bR_{l}^{(i)} (x_l)$ is $N \times N$,
the complexity might be prohibitively high if $N$ is large.
However, from 
\eqref{EQ:iRl}, 
we can see that
\be
\bR_{l+1}^{(i)} (x_{l+1})
= \bh_{l+1} \bh_{l+1}^\rH (x_{l+1} - q_{l+1}^{(i-1)})
+ \bR_{l}^{(i)} (q_l^{(i)}).
	\label{EQ:Rxq}
\ee
Thus, using the matrix inversion lemma \cite{Harv97},
we have
\begin{align}
& \bR_{l+1}^{(i)} (x_{l+1})^{-1}
= \bR_{l}^{(i)} (q_l^{(i)})^{-1} \cr
& \quad - \nu_{l+1}^{(i)} (x_{l+1})
\bR_{l}^{(i)} (q_l^{(i)})^{-1} \bh_{l+1}
\bh_{l+1}^\rH \bR_{l}^{(i)} (q_l^{(i)})^{-1} , 
	\label{EQ:milR}
\end{align}
for $x_{l+1} \in \{0,1\}$, where
$$
\nu_{l+1}^{(i)} (x_{l+1})
=\frac{x_{l+1} - q_{l+1}^{(i-1)}
}{1+(x_{l+1} - q_{l+1}^{(i-1)})
\bh_{l+1}^\rH \bR_{l}^{(i)} (q_l^{(i)})^{-1} \bh_{l+1}}.
$$
Provided that $\bR_{l}^{(i)} (q_l^{(i)})^{-1}$ is available,
according to \eqref{EQ:milR},
the complexity to find 
$\bR_{l+1}^{(i)} (x_{l+1})^{-1}$ is 
mainly due to 
$\bR_{l}^{(i)} (q_l^{(i)})^{-1} \bh_{l+1}$, which becomes
$O(N^2)$ in terms of the number of multiplications.
In addition, per iteration, 
there might be $L$ updates for $\{q_l\}$. Thus,
the computational complexity per iteration becomes
$O(L N^2)$.

Note that using the matrix determinant lemma \cite{Harv97},
from \eqref{EQ:Rxq},
the term $ \ln \det(\bR_l^{(i)} (x_l))$  in \eqref{EQ:chil}
can be updated as follows:
\begin{align}
&\ln \det(\bR_{l+1}^{(i)} (x_{l+1}))
= \ln \det(\bR_{l}^{(i)} (q_{l}^{(i)})) \cr
& \quad +
\ln (1+ (x_{l+1} - q_{l+1}^{(i-1)})
\bh_{l+1}^\rH \bR_{l}^{(i)} (q_{l}^{(i)})^{-1} 
\bh_{l+1}).
\end{align}
Since $\bR_{l}^{(i)} (q_{l}^{(i)})^{-1} 
\bh_{l+1}$ is available in finding 
$\bR_{l+1}^{(i)} (x_{l+1})^{-1}$ as shown in \eqref{EQ:milR},
there is no significant additional 
computational complexity to 
update the term $ \ln \det(\bR_l^{(i)} (x_l))$.
Thus, the computational complexity 
per iteration remains $O(LN^2)$.

In summary,
the computational complexity 
of the CAVI algorithm is independent of the number of
active transmit antennas, $K$,
while it grows linearly with 
the number of transmit antennas, $L$, and 
quadratically with the number of receive antennas, $N$.
Note that the complexity of the ML approach in \eqref{EQ:ml_ga}
or the approach in \cite{Lin15, Freud18} 
is $O(|\cX_K|) = O(\binom{L}{K})$, which grows exponentially
with $K$ (for a fixed $L$). 
In particular, when $K \ll L$, we can show that
$\binom{L}{K} \approx \frac{L^K}{K!}$.
Although the complexity for a given $\bx$ can be an order of
$O(K^3)$ as shown \cite{Freud18},
the overall complexity is $O(K^3 L^K)$ for $K \ll L$.
Thus, if $K \ge 3$, the complexity of the ML approach 
in \cite{Lin15,Freud18} or that in \eqref{EQ:ml_ga}
is much higher than that of the CAVI algorithm\footnote{As shown in
Section~\ref{S:Sim}, the CAVI algorithm requires several
iterations to converge (less than 10 iterations). Thus,
its complexity is $O(N_{\rm run} LN^2)$, where $N_{\rm run} < 10$.
Here, $N_{\rm run}$ represents the number of iterations.}
in this paper,
while the performance of the CAVI algorithm is worse
than that in \cite{Lin15, Freud18}, because it provides
an approximate solution to the ML problem in \eqref{EQ:ml_ga}
(or that in \cite{Lin15, Freud18}).


\section{Performance Analysis}	\label{S:PA}

In this section, we consider the performance of the ML
detection in \eqref{EQ:ml_ga} based on the pairwise error probability (PEP)
analysis, which can result in an approximate 
probability of index error,
i.e., the probability that the estimated set of active
indices is not the same as the true set.
Based on the analysis in this section,
we are able not only to show that
the solution to the approximate ML formulation in \eqref{EQ:ml_ga} 
can achieve a full coding gain, but also to compare
the performance of the CAVI algorithm with the ML performance
(which will be shown in Section~\ref{S:Sim}).

For tractable analysis,
we consider the following assumptions.
\begin{itemize}
\item[{\bf A1)}] For DM symbols, 
a constant modulus modulation scheme (e.g.,
4-QAM) is employed.
i.e., $|s_{m,l}| = 1$, $l \in \cI$.
\item[{\bf A2)}] The elements of $\bH$ are 
independent and identically distributed (i.i.d.)
with zero-mean and finite variance. In particular, we assume that
$H_{n,l} \sim \cC \cN(0, 1/N)$.
\end{itemize}

Based on the PEP analysis \cite{BiglieriBook},
the probability of index error is upper-bounded as
\begin{align}
P_{\rm ie} & 
= \Pr(\hat \cI \ne \cI) 
= \Pr(\hat \bx \ne \bx) \cr
& \le \sum_{\bx \in \cX_K} 
\sum_{\bx^\prime \in \cX_K, \bx^\prime \ne \bx}
P(\bx\to \bx^\prime) \Pr(\bx) \cr
& = \frac{1}{|\cX_K|} \sum_{\bx \in \cX_K} 
\sum_{\bx^\prime \in \cX_K, \bx^\prime \ne \bx}
P(\bx\to \bx^\prime),
	\label{EQ:Pie}
\end{align}
where $\hat \cI$ represents the estimated index set and
$P(\bx\to \bx^\prime)$ is the 
PEP, i.e., the probability that
$\bx^\prime$ is decoded when $\bx$ is the correct index vector
(assuming that there are only two possible index vectors,
$\bx$ and $\bx^\prime$).
To find the PEP,
let 
$\by = [\by_1^\rH \ \ldots \ \by_M^\rH]^\rH$,
$\bDelta = \bV(\bx) - \bV(\bx^\prime)$,
and
$\cL (\bx) = \ln \prod_{m=1}^M f(\by_m\,|\,\bx)$.
When the ML detection in \eqref{EQ:ml_ga} is considered,
the PEP is given by
\begin{align}
P(\bx \to \bx^\prime) & = 
  \Pr( \cL (\bx) < \cL (\bx^\prime) \,|\, \bx) \cr
& = \Pr\left(\sum_m \by_m^\rH \bDelta \by_m >  
d(\bx, \bx^\prime)
 \,|\, \bx \right) ,
	\label{EQ:PEP}
\end{align}
where
\begin{align}
d(\bx, \bx^\prime) = M (\phi (\bx^\prime)-\phi (\bx)).
	\label{EQ:dxx}
\end{align}

For tractable analysis in finding the probability of index error,
we only consider the case that the Hamming distance 
between $\bx$ and $\bx^\prime$ is a minimum\footnote{In the high
SNR regime, 
the PEP associated with a Hamming distance
larger than the minimum Hamming distance
becomes negligible compared
to the PEP associated with the minimum Hamming distance.
Thus, 
a good approximate probability of index error
can be obtained by considering only the PEPs 
associated with the minimum Hamming distance.}.
Since $|{\rm supp}(\bx)| = |{\rm supp}(\bx^\prime)| = K$,
the minimum
Hamming distance between $\bx$ and $\bx^\prime$ is 2 and
\be
{\rm supp}(\bee) = \{l_1, l_2\},
	\label{EQ:sbee1}
\ee
where $\bee = \bx^\prime - \bx$ and
$l_1 \ne l_2$, $l_1, l_2 \in \{1,\ldots, L\}$.
For convenience, we assume that 
\begin{align}
l_1 = 1 \in {\rm supp}(\bx) \ \mbox{and} \
l_2 = 2 \in {\rm supp}(\bx^\prime).
	\label{EQ:sbee2}
\end{align}
For given $\bx$, let
$\bar \cX(\bx) = \{\bx^\prime\,|\, 
d_H(\bx^\prime, \bx) = 2, \ \bx^\prime \in \cX_K\}$,
where $d_H(\bx^\prime,\bx)$ represents the
Hamming distance between $\bx^\prime$ and $\bx$.
Since the PEP associated with the two index vectors that
have the Hamming distance larger than the minimum might be
sufficiently low, the following approximation might be reasonable:
$\sum_{\bx^\prime \in \cX_K, \bx^\prime \ne \bx}
P(\bx\to \bx^\prime)
\approx 
\sum_{\bx^\prime \in \bar \cX (\bx)} P(\bx\to \bx^\prime)$,
and from \eqref{EQ:Pie}, we can have the following approximate
probability of index error:
\be
P_{\rm ie} \approx
\frac{1}{|\cX_K|} \sum_{\bx \in \cX_K} 
\sum_{\bx^\prime \in \bar \cX(\bx)} 
P(\bx\to \bx^\prime).
	\label{EQ:Pie2}
\ee

In general, since the conditional 
probability density function
(pdf) of $\by_m$ 
for given $\bx$ is a Gaussian mixture as shown in \eqref{EQ:lhx}, 
it is difficult to obtain 
a closed-form expression for the PEP. Thus, 
with $l_1 = 1$,
we may need to consider the following Gaussian approximation for the
interference-plus-noise:
\be
\by_m = \bh_1 s_{m,1} + \bee_m,
\ s_{m,1} \in \cS, \ m = 1,\ldots, M,
	\label{EQ:ym1}
\ee
where $\bee_m \sim \cC \cN(\b0, \bD)$, which is the sum
of the other active DM signals and the background noise. 
Here,
$\bD = \bar \bH \bar \bH^\rH + \gamma^{-1} \bI$.
The approximation in \eqref{EQ:ym1} can be reasonable
if the sum of active DM signals is well approximated by a Gaussian
random variable or the noise term in $\bee_m$ is dominant.

\begin{mylemma}	\label{L:PEP}
Consider $\bx$ and $\bx^\prime$ with a Hamming distance of 2.
With \eqref{EQ:sbee1} and \eqref{EQ:sbee2},
under the assumption of {\bf A1},
let
$\bV(\bx)^{-1} = \bD + \bh_{1} \bh_{1}^\rH$ and
$\bV(\bx^\prime)^{-1} = \bD + \bh_{2} \bh_{2}^\rH$.
Here, $\bar \bH$ is the submatrix of $\bH$ with the column
vectors corresponding to the indices of 
${\rm supp}(\bx) \cap {\rm supp}(\bx^\prime)$.
Then, for given $\bx$, 
based on the Gaussian approximation for the
interference-plus-noise in \eqref{EQ:ym1},
an upper-bound on the conditional PEP
is given by
\begin{align}
& \Pr\left(\sum_{m=1}^M \by_m^\rH \bDelta \by_m  > d\,|\, 
\bh_1, \bh_2, \bD \right) \cr
& \quad \le e^{-\lambda d}
\left( \kappa (\lambda) \exp \left(-
\gamma_{\rm pep} (\lambda)
\right) \right)^M,
	\label{EQ:y_chi}
\end{align}
where $\lambda \ge 0$ and
\begin{align} 
\kappa(\lambda) & =  
\frac{(\alpha_1 +1)(\alpha_2 + 1) \theta_2 (\lambda)}{
\lambda ( \alpha_1 + 1 +\lambda(
\alpha_1 + \theta_2(\lambda)|\beta|^2))} \cr
\gamma_{\rm pep} (\lambda) & = 
\alpha_1 - 
\frac{(\alpha_1 +1) (\alpha_1 + \theta_2(\lambda) |\beta|^2)}{
\alpha_1 +1+ \lambda (\alpha_1 + \theta_2(\lambda) |\beta|^2)}.
	\label{EQ:alal}
\end{align}
Here,
$\gamma_{\rm pep} (\lambda)$ represents the PEP-SNR
(i.e., the effective SNR for PEP),
and
$\alpha_l = \bh_l^\rH \bD^{-1} \bh_l$,
$\beta =  \bh_1^\rH \bD^{-1} \bh_2$, and
$\theta_l (\lambda) = \frac{\lambda}{1 + \alpha_l - \alpha_l \lambda}$.
\end{mylemma}
\begin{IEEEproof}
See Appendix~\ref{A:PEP}.
\end{IEEEproof}
In \eqref{EQ:alal}, we can 
observe that the coding
gain in detecting $\cI$ is up to $M$ due to $M$ independent
data symbols, $\{s_{m,1}\}$. To see this clearly,
we consider an example with $K = 1$ (i.e., only one antenna is active).
In this case, $\by_m$ becomes
$$
\by_m = \bh_l s_{m,l} + \bn_m, \ l \in \{1,\ldots, L\},
\ m = 1,\ldots, M,
$$
where $\bh_l \in \cH = \{\bh_1, \ldots, \bh_L\}$ becomes the signal
and the $s_{m,l}$'s are $M$ independent random weights 
(like the fading coefficients) multiplied to
the signal, $\bh_l$. Here, $\cH$ can be seen as the signal constellation.
Thus, in detecting $\bh_l \in \cH$, the coding
gain can be up to $M$ as shown in \eqref{EQ:alal}.

With \eqref{EQ:sbee1}
and \eqref{EQ:sbee2}, since
\be
\phi(\bx) = \ln\det( \bD
+ \bh_{1} \bh_{1}^\rH ) \ \mbox{and} \
\phi(\bx^\prime) = \ln\det( \bD + \bh_{2} \bh_{2}^\rH ) ,
\ee
from \eqref{EQ:dxx}, we have 
\begin{align}
d(\bx, \bx^\prime) 
= M \ln \frac{1+\alpha_{2}}{1+\alpha_{1}}.
	\label{EQ:d21}
\end{align}
Thus, with $d  =
 M \ln \frac{1+\alpha_{2}}{1+\alpha_{1}}$,
the (average) PEP in \eqref{EQ:PEP} can be obtained by taking
the expectation of the conditional PEP in
\eqref{EQ:alal} with respect to
$\bh_1$, $\bh_2$, and $\bD$.
Unfortunately, 
this expectation cannot be carried out unless
the joint pdf of $\alpha_l$ and $\beta$ is available.
Even if the joint pdf is available,
a closed-form expression for the expectation may not be easy to find.
Thus, we may only consider asymptotic cases.

As in \cite{Tse99} and \cite{TulinoBook},
for a large $N$ with fixed $\eta$, where $\eta = \frac{Q}{N}$,
we expect that $\alpha_l$ converges to a constant,
i.e.,
\be
\alpha_l \to \bar \alpha = \gamma_{\rm mmse} (\gamma, \eta),
	\label{EQ:balpha}
\ee
where 
$\gamma_{\rm mmse} (\gamma, \eta)$ is the asymptotic
SINR of the minimum mean squared error (MMSE) receiver.
Under the assumption of {\bf A2},
a closed form expression for
$\gamma_{\rm mmse} (\gamma, \eta)$ 
can be found in \cite[Eq. (9)]{Tse99}.
In this case, 
we have $d(\bx, \bx^\prime) = 0$ from \eqref{EQ:d21}.

\begin{mylemma}	\label{L:Asym}
For a sufficiently large $N$ with fixed $\eta$,
the asymptotic PEP-SNR is approximated  as follows:
\begin{align}
\gamma_{\rm pep} (\lambda)
\approx {\tilde \gamma}_{\rm pep} (\lambda) 
= \bar \alpha - 
\frac{(\bar \alpha +1)
\left(\bar \alpha + \frac{\bar \omega \bar \theta_2(\lambda)}{N} \right)}{
\bar \alpha +1+ \lambda \left( \bar \alpha + 
\frac{\bar \omega \bar \theta_2(\lambda)}{N} \right)},
	\label{EQ:ub_SNR}
\end{align}
where 
\begin{align*}
\bar \omega = 
\frac{\gamma_{\rm mmse} ( \gamma (2 + \gamma), \eta )}{2 \gamma^{-1}+1}
\ \mbox{and} \ 
\bar \theta_l (\lambda) = \frac{\lambda}{1 + \bar \alpha - \bar
\alpha \lambda}.
\end{align*}
\end{mylemma}
\begin{IEEEproof}
See Appendix~\ref{A:Asym}.
\end{IEEEproof}

For a large $N$ with fixed $\eta$,
the asymptotic PEP-SNR and
the asymptotic $\kappa (\lambda)$,
which is denoted by $\tilde \kappa (\lambda)$,
are not dependent on the realizations
of $\bh_1$, $\bh_2$, and $\bD$.
Thus, the asymptotic conditional PEP becomes
the asymptotic PEP,
which can be approximated as follows from \eqref{EQ:y_chi}:
\be
P(\bx \to \bx^\prime) \approx 
\tilde P_2 (\lambda) =
\left( \tilde \kappa(\lambda) \exp\left(
- \tilde \gamma_{\rm pep} (\lambda)
\right) \right)^M.
	\label{EQ:P2}
\ee
The parameter $\lambda$
can be optimized for a tight bound as follows:
\begin{align}
\lambda^* = \argmax_{0 < \lambda < \bar \lambda}
{\tilde \gamma}_{\rm pep} (\lambda) - \ln \tilde \kappa (\lambda),
\end{align}
where
$\bar \lambda = \frac{\bar \alpha +1}{\bar \alpha}$.
Thus, $\tilde P_2 (\lambda^*)$ is a tight approximate PEP.
Finally, 
since $\tilde P_2 (\lambda^*)$ does not depend on
$\bx$ and $\bx^\prime$ and $|\bar \cX(\bx)| = L-Q$, 
from \eqref{EQ:Pie2} and \eqref{EQ:P2},
an approximate probability of index error 
can be given by
\begin{align}
P_{\rm ie} 
& \approx (L-Q) \tilde P_2(\lambda^*) \cr
& =(L-Q) 
\tilde \kappa(\lambda^*)^M \exp\left(
- M \tilde \gamma_{\rm pep} (\lambda^*)
\right).
	\label{EQ:APim}
\end{align}
Consequently,
in \eqref{EQ:APim}, we can observe that
although \eqref{EQ:ml_ga} 
is an approximate ML formulation for the index detection
(based on the Gaussian approximation),
its solution can achieve a full coding gain
in RCSM.

Note that \eqref{EQ:APim} is an approximation
and would be reasonable for a large system
(i.e., a large $N$ with fixed $\eta$).
If $N$ is not sufficiently large, the expectation of
the conditional PEP in \eqref{EQ:y_chi}
with respect to $\bh_1$, $\bh_2$, and $\bD$ is to be carried out 
to obtain the average PEP. Thus, 
\eqref{EQ:APim} may not be reasonable for a small system
(i.e., when $N$ is small).
In particular, as shown in 
\eqref{EQ:y_chi}, the conditional probability of index error
is proportional to $e^{-M \gamma_{\rm pep}}$.
Thus, the average PEP becomes 
proportional to $\uE[e^{-M \gamma_{\rm pep}}]$,
while we consider $e^{-M \uE[\gamma_{\rm pep}]}$ in
\eqref{EQ:APim}, which might be correct in the asymptotic
case where $N \to \infty$ with fixed $\eta$ as mentioned earlier. However,
for a finite $N$, \eqref{EQ:APim} leads to 
an underestimate for a high SNR (with a finite $N$)
due to Jensen's inequality.

\section{Simulation Results}	\label{S:Sim}

\begin{figure}[thb]
\begin{center}
\includegraphics[width=\figwidth]{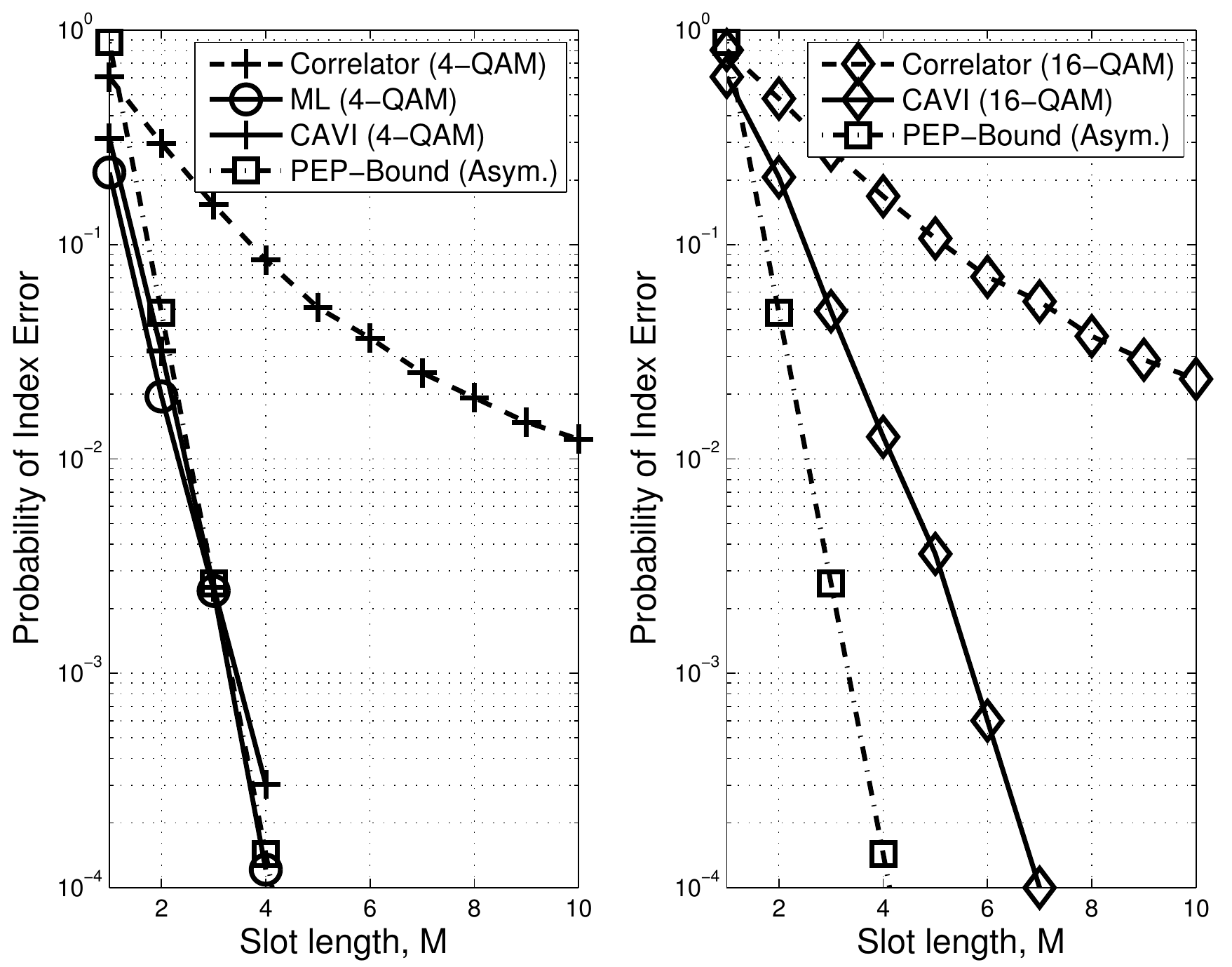} \\
\hskip 2.5cm (a) \hskip 3.5cm (b) 
\end{center}
\caption{The probabilities of index error 
of the correlator based detector, the ML detector 
(with Gaussian approximation in \eqref{EQ:ml_ga}),
and CAVI algorithm with $(\mu, N_{\rm run}) = (0.5, 12)$
for various slot lengths, $M$,
when $N = 40$, $L = 20$, $K = 4$, and SNR $= 10$ dB:
(a) 4-QAM; (b) 16-QAM.}
        \label{Fig:plt4}
\end{figure}

In this section, we present simulation results
to see the performance of the CAVI algorithm
for the index detection in RCSM.
For simulations, we consider the assumptions
of {\bf A1} and {\bf A2} with $4$-QAM
(except for Fig.~\ref{Fig:plt4} (b)).


For comparisons, 
we consider the ML performance
(i.e., the performance of the ML detection with
Gaussian approximation in \eqref{EQ:ml_ga})
using simulation results as well as 
the approximate probability
of index error in \eqref{EQ:APim}.
We expect that the CAVI algorithm can provide a near ML performance so
that it can be used as a low-complexity approach to find an approximate
ML solution. By comparing the ML performance
with 
the approximate probability
of index error in \eqref{EQ:APim}, we can also confirm that
the coding gain of RCSM, which is $M$.
In addition, we consider
the correlator based detector as a (very) low-complexity
solution.

\begin{figure}[thb]
\begin{center}
\includegraphics[width=\figwidth]{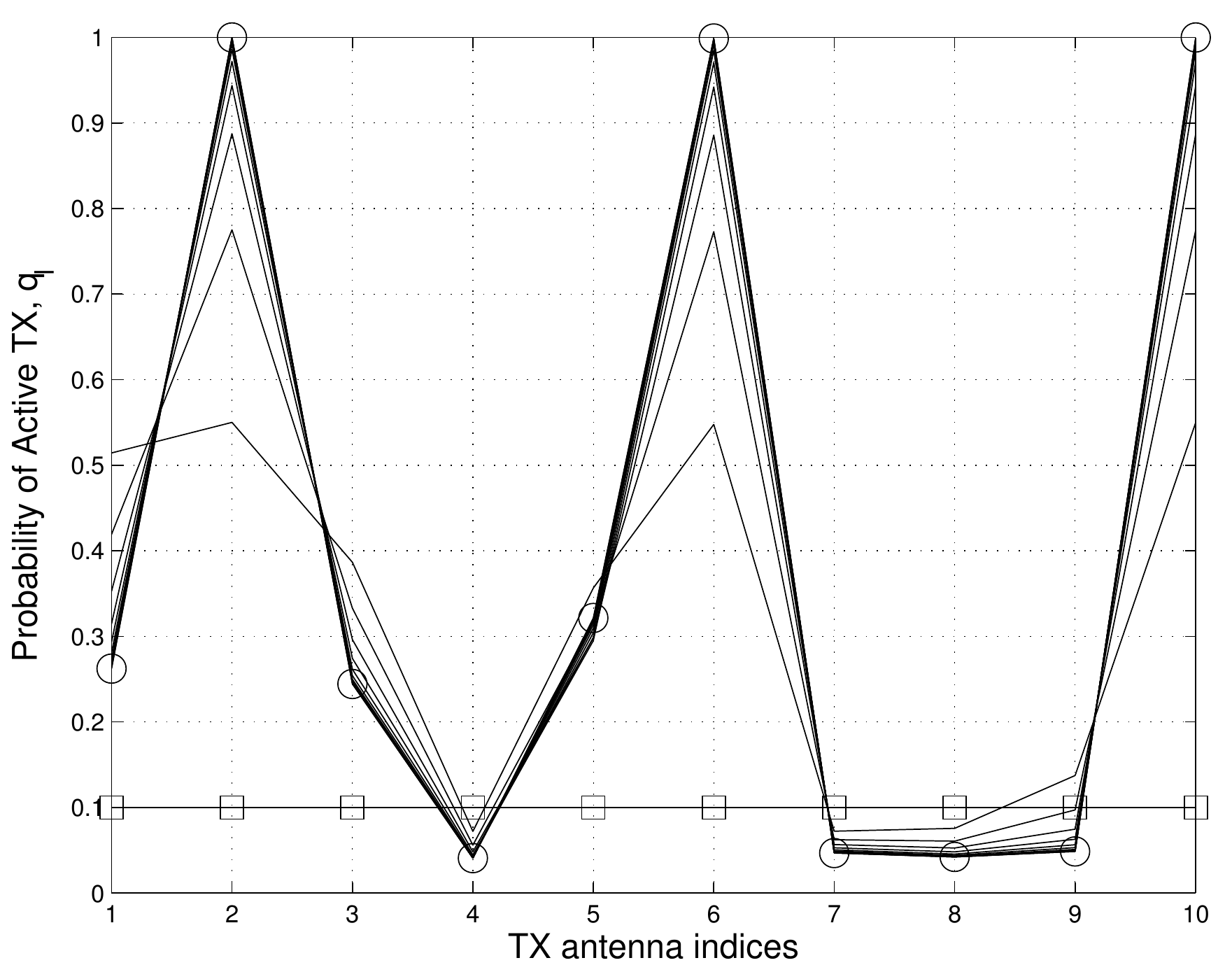}
\end{center}
\caption{The convergence behavior of $q_l^{(i)}$ for the CAVI algorithm
when $N = L = 10$, $K = 3$, $M = 2$, $\gamma = 10$ dB,
and $(\mu, N_{\rm run}) = (0.5, 12)$. The initial
variational probabilities of $q_l^{(0)}$ are set to $1/L$, 
which are represented by $\Box$ markers.
The final 
variational probabilities are represented by $\circ$ markers.}
        \label{Fig:plt_conv}
\end{figure}

Since the CAVI algorithm is an iterative algorithm,
its performance depends on the number of iterations.
In Fig.~\ref{Fig:plt_conv},
we present the variational probability, $q_l^{(i)}$,
for each iteration
when $N = L = 10$, $K = 3$, $M = 2$, $\gamma = 10$ dB,
and $(\mu, N_{\rm run}) = (0.5, 12)$.
The initial probabilities of $q_l^{(0)}$ are set to $1/L$.
The set of active indices 
is $\cI = \{2, 6, 10\}$.
It is shown that the CAVI algorithm can provide
large variational probabilities for $l \in \cI$
after a certain number of iterations.
As a result, we can expect the correct result
for the index detection after 
convergence.

\begin{figure}[thb]
\begin{center}
\includegraphics[width=\figwidth]{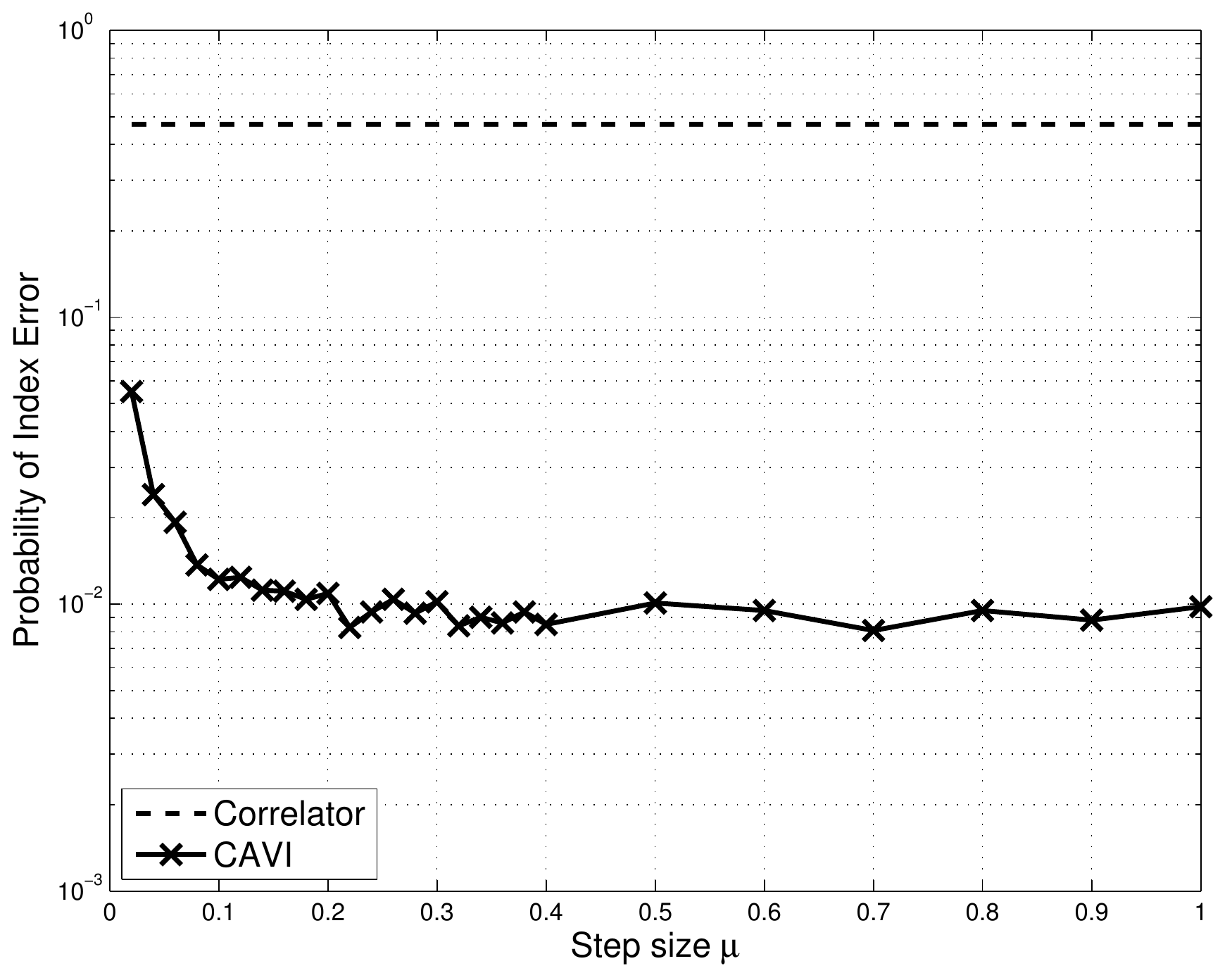}
\end{center}
\caption{The probability of index error of the CAVI algorithm
as a function of the step size, $\mu$,
when $N = 10$, $L = 20$, $K = 2$, $M = 4$, $\gamma = 10$ dB,
and $N_{\rm run} = 10$.}
        \label{Fig:plt1}
\end{figure}

Fig.~\ref{Fig:plt1} shows the probability of index
error as a function of the step-size, $\mu$,
when $N = 10$, $L = 20$, $K = 2$, $M = 4$, $\gamma = 10$ dB,
and $N_{\rm run} = 10$.
We can see that the performance is improved by increasing
$\mu$. However, when $\mu \ge 0.4$, there is no performance
improvement.
We also show the performance of the correlator based detector
with the dashed line in Fig.~\ref{Fig:plt1}.
Clearly, the CAVI algorithm can perform better than
the correlator based detector as it can provide an approximate
ML solution (at the cost of a higher computational complexity
than that of the correlator based detector).

\begin{figure}[thb]
\begin{center}
\includegraphics[width=\figwidth]{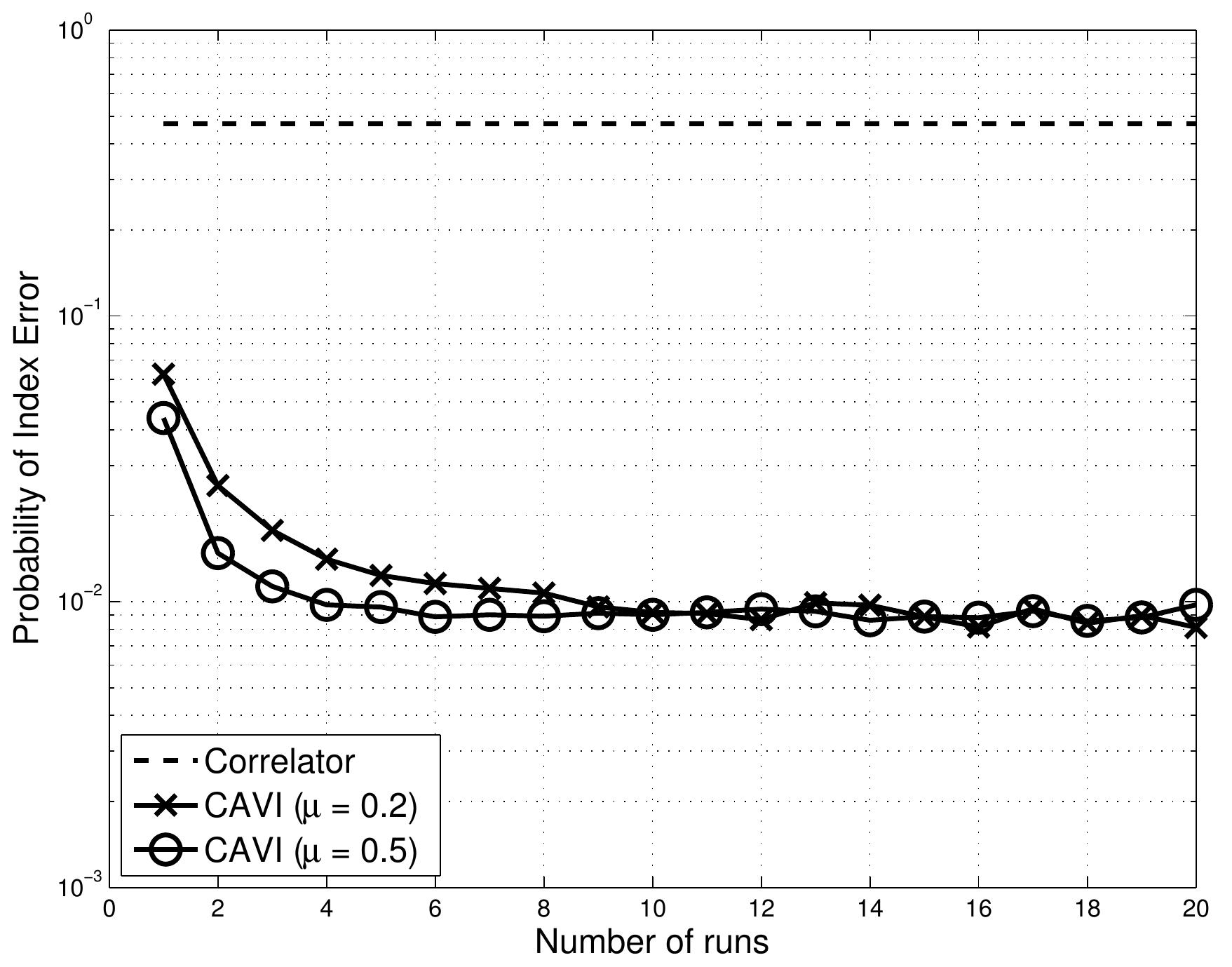}
\end{center}
\caption{The probability of index error of the CAVI algorithm
as a function of the number of runs,
when $N = 10$, $L = 20$, $K = 2$, $M = 4$, $\gamma = 10$ dB,
and $\mu \in \{0.2, 0.5\}$.}
        \label{Fig:plt2}
\end{figure}

The impact of the number of iterations, $N_{\rm run}$,
on the performance is shown in
Fig.~\ref{Fig:plt2}
when $N = 10$, $L = 20$, $K = 2$, $M = 4$, $\gamma = 10$ dB,
and $\mu \in \{0.2, 0.5\}$.
It is clearly shown that the performance can be improved
by more iterations, while 10 iterations (i.e., $N_{\rm run} = 10$)
might be sufficient for convergence.

\begin{figure}[thb]
\begin{center}
\includegraphics[width=\figwidth]{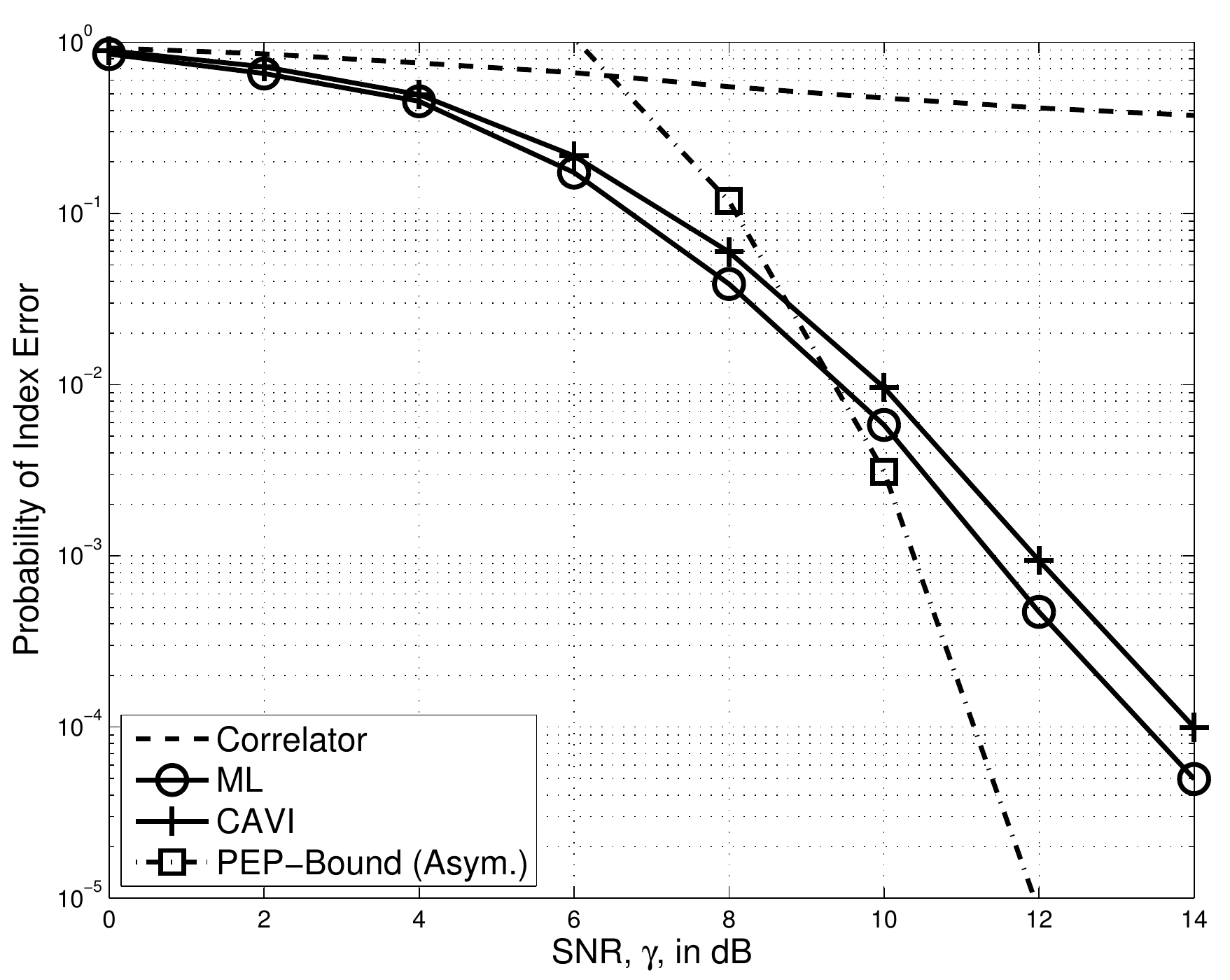} \\
(a) \\
\includegraphics[width=\figwidth]{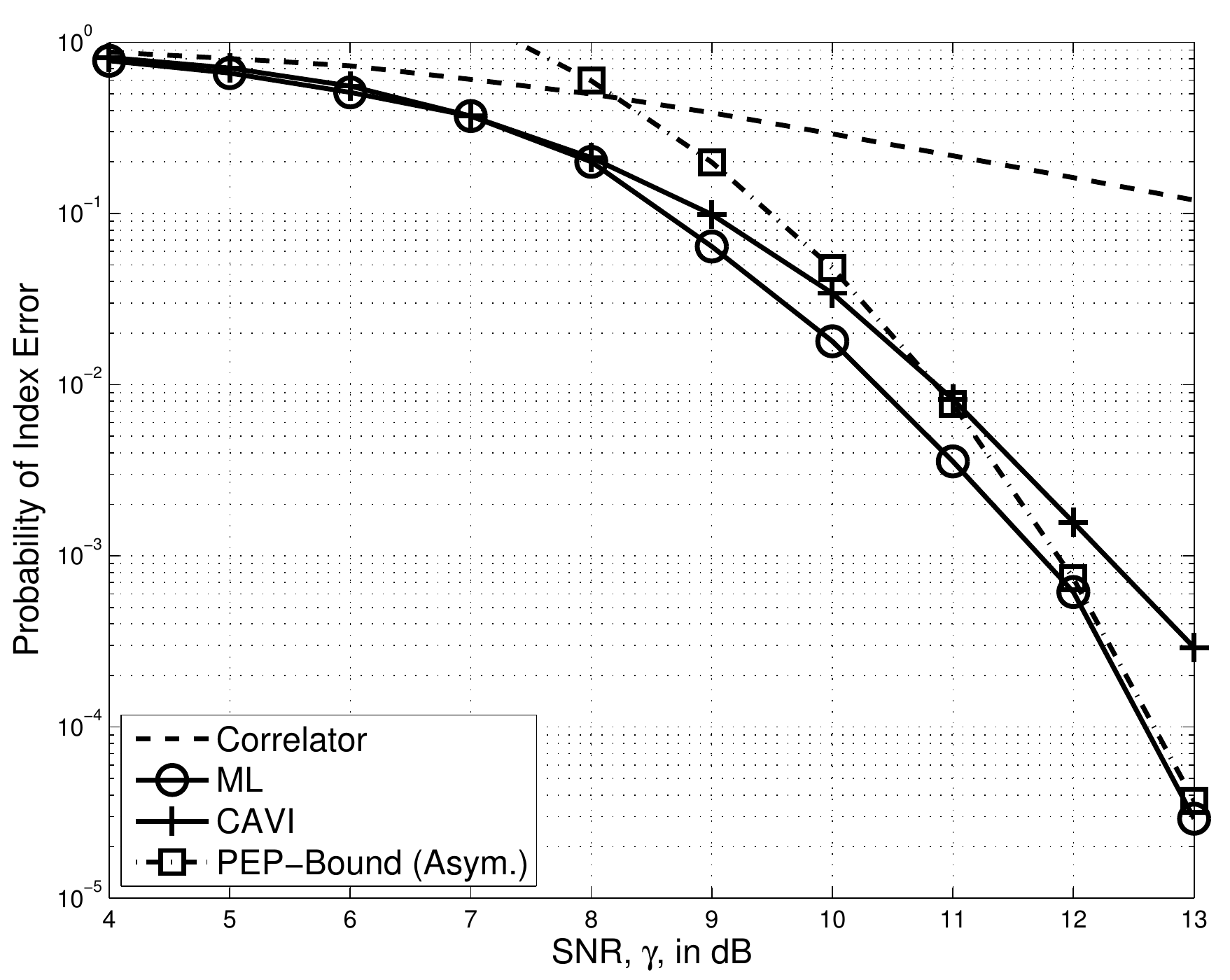} \\
(b) \\
\end{center}
\caption{The probabilities of index error 
of the correlator based detector, ML detector 
(with Gaussian approximation in \eqref{EQ:ml_ga}),
and CAVI algorithm with $(\mu, N_{\rm run}) = (0.5, 12)$
for various values of SNRs:
(a) $N = 10$, $L = 20$, $K = 2$, and $M = 4$;
(b) $N = 40$, $L = 20$, $K = 4$, and $M = 2$.}
        \label{Fig:plt3}
\end{figure}

The performances of the 
correlator based detector, the ML detector with
Gaussian approximation \cite{Lin15} (i.e.,
\eqref{EQ:ml_ga}), and
the CAVI algorithm are shown for different values of SNR
in Fig.~\ref{Fig:plt3}.
Since the ML detector
and CAVI algorithm are to jointly detect active 
indices, it does not significantly suffer from the interference
from the signals from the other active transmit
antennas (at a high SNR). 
On the other hand, in the correlator based detector,
although the noise is negligible (at a high SNR),
its performance is still degraded by the interference.
Thus, as shown in Fig.~\ref{Fig:plt3},
the performance gap between the CAVI algorithm 
and the correlator based detector becomes wider 
as the SNR increases, 
while the performance of the CAVI algorithm
is slightly degraded from that of the ML detector in \eqref{EQ:ml_ga}.
We can also see that \eqref{EQ:APim}
might be reasonable for a large system (as shown in
Fig.~\ref{Fig:plt3} (b)),
while at a high SNR
\eqref{EQ:APim} becomes an underestimate
of the probability of index error
as mentioned earlier.

As shown in \eqref{EQ:APim},
the slot length, $M$, becomes the coding gain.
Thus, we expect that the probability of index error decreases
exponentially with $M$, which can be confirmed by
the simulation results in Fig.~\ref{Fig:plt4} (a).
That is, we can see that
the probability of index error 
of the ML detector and the asymptotic PEP-bound
in \eqref{EQ:APim}
(that are shown by the solid line with $\circ$ markers
and the dashed line with $\Box$ markers, respectively)
agree with each other in Fig.~\ref{Fig:plt4} (a). 
We can also observe that 
the performance of the CAVI algorithm is slightly 
worse than that of the ML detector,
which means the CAVI algorithm can have a near ML performance
(of the detection in \eqref{EQ:ml_ga}).

In Fig.~\ref{Fig:plt4} (b), we also present simulation results with
16-QAM. Compared with the performances with 4-QAM,
the performances with 16-QAM are degraded in both the correlator
based detector and the CAVI algorithm.
From this, we expect to have a trade-off between
the spectral efficiency and the detection
performance (in terms of the probability of index error).
Since the performance analysis in Section~\ref{S:PA}
is carried out under the assumption of {\bf A1} (e.g., 4-QAM),
we need to extend the analysis with non-constant modulus modulation
schemes to clearly quantify the performance
degradation by 16-QAM, which might be a further research topic in the future.

\begin{figure}[thb]
\begin{center}
\includegraphics[width=\figwidth]{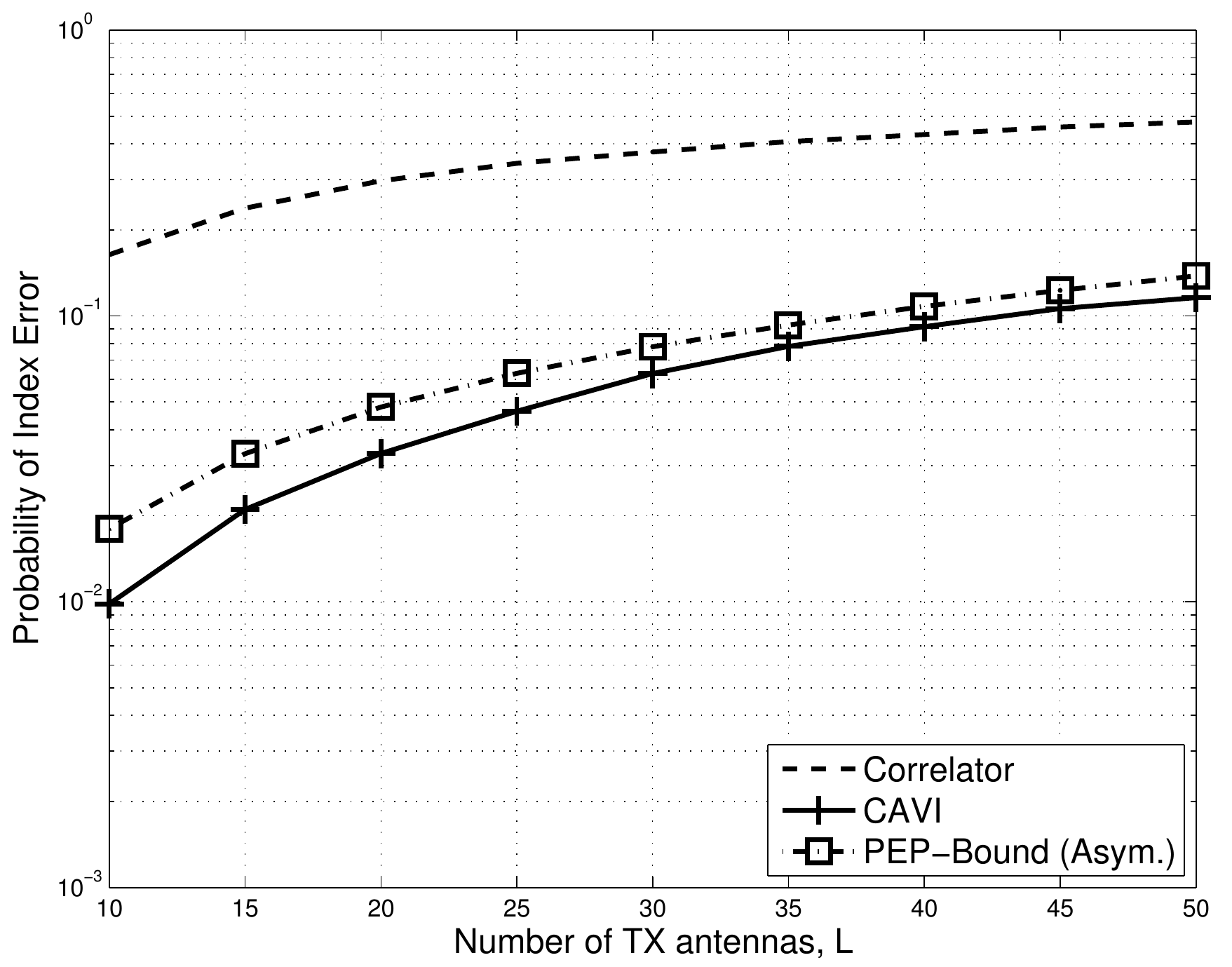} \\
(a) \\
\includegraphics[width=\figwidth]{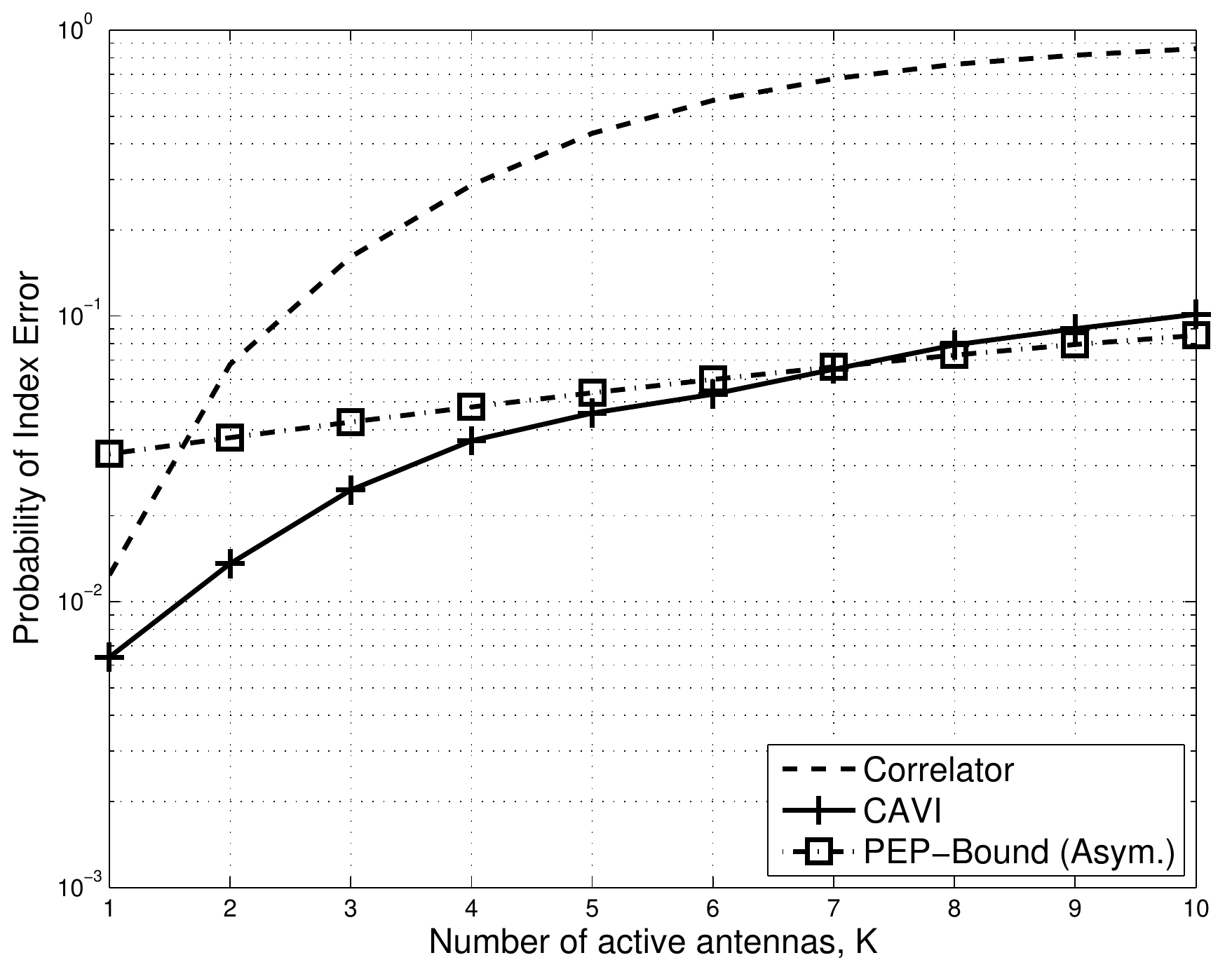} \\
(b) 
\end{center}
\caption{Performances
of the correlator based detector
and CAVI algorithm with $(\mu, N_{\rm run}) = (0.5, 12)$
when $N = 40$, $M = 2$, and SNR $= 10$ dB:
(a) the probabilities of index error  as functions of
different values of transmit antennas, $L$ (with $K = 4$);
(b) the probabilities of index error  as functions of
different values of active transmit antennas, $K$ (with $L = 20$).}
        \label{Fig:plt5}
\end{figure}

Fig.~\ref{Fig:plt5} (a) shows the 
the probabilities of index error 
of the correlator based detector and CAVI algorithm
for different values of transmit antennas, $L$,
where we can see that 
the probability of index error increases with $L$.
We also see that
the probability of index error increases with $K$
in Fig.~\ref{Fig:plt5} (b).
Clearly, from Fig.~\ref{Fig:plt5} (a) and (b), we see 
a trade-off between the spectral efficiency and the detection
performance via $L$ or $K$, because 
the number of IM bits increases with both $L$ and $K$
(as shown in \eqref{EQ:Bsym}).

\begin{figure}[thb]
\begin{center}
\includegraphics[width=\figwidth]{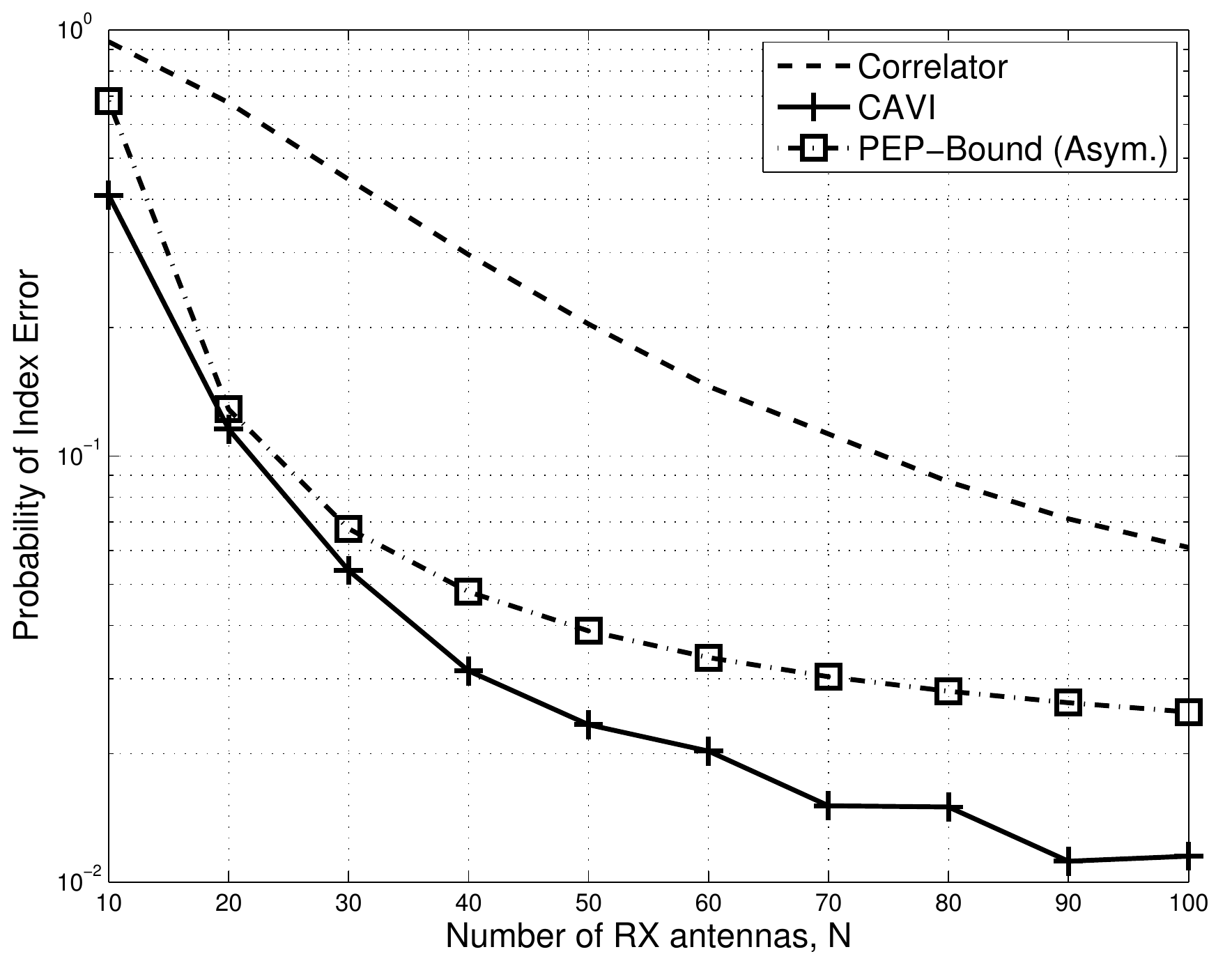}
\end{center}
\caption{The probabilities of index error 
of the correlator based detector
and CAVI algorithm with $(\mu, N_{\rm run}) = (0.5, 12)$
for different values of receive antennas, $N$,
when $L = 20$, $K = 4$, $M = 2$, and SNR $= 10$ dB.}
        \label{Fig:plt7}
\end{figure}

In general, we expect that the performance of index detection 
can be improved by increasing
the number of receive antennas, $N$,
which can be confirmed by the simulation results in
Fig.~\ref{Fig:plt7}. It is interesting to note that the performance
gap between the CAVI algorithm and 
the correlator based detector
becomes narrower as $N$ increases.
For fixed $L$ and $K$, when $N$ increases,
the interference can be better suppressed in
the correlator based detector.
However, if $N$ is not too large, it might be necessary
to carefully deal with the interference and any joint detection
approach (e.g., the CAVI algorithm) might be required for a reasonable
performance.

\begin{figure}[thb]
\begin{center}
\includegraphics[width=\figwidth]{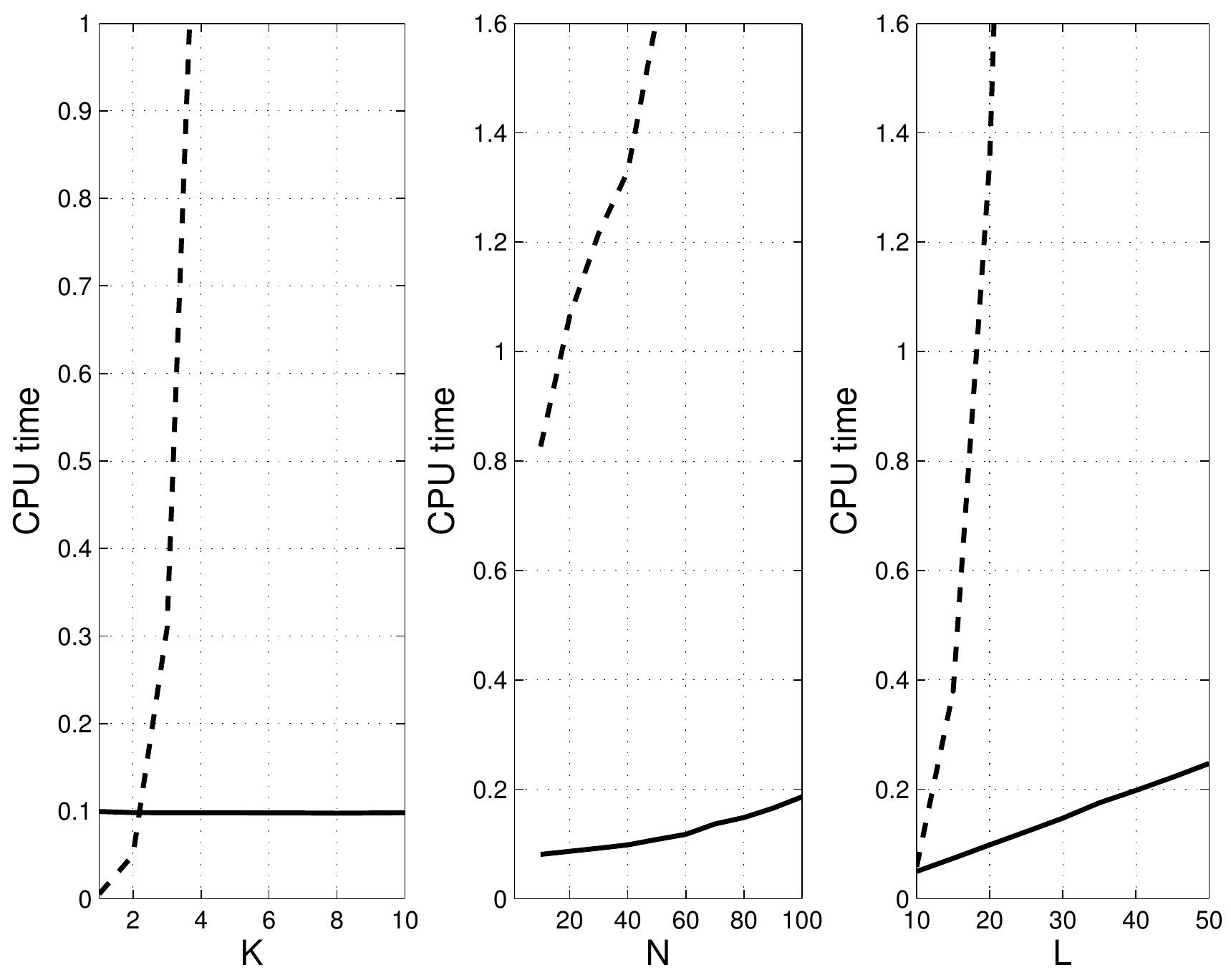}
\end{center}
\caption{The computational complexity
of the CAVI algorithm (shown by the solid lines)
and the ML detector in \eqref{EQ:ml_ga}
(shown by the dashed lines)
as functions of key parameters
when SNR $= 10$ dB and $(\mu, N_{\rm run}) = (0.5, 10)$ for the
CAVI algorithm:
(a) the number of active transmit antennas, $K$ 
(with $N = 40$ and $L = 20$);
(b) the number of receive antennas, $N$
(with $K = 4$ and $L = 20$);
(c) the number of transmit antennas, $L$
(with $N = 40$ and $K = 4$).}
        \label{Fig:plt_times}
\end{figure}

The computational complexity of the CAVI algorithm
was discussed in Subsection~\ref{SS:LC},
where we show that the complexity order is $O(N_{\rm run} N^2 L)$.
The computational complexity
of the ML detection in \eqref{EQ:ml_ga} is also shown for comparisons.
As mentioned earlier,
the ML detection in \eqref{EQ:ml_ga} with $M = 1$ is considered
in \cite{Lin15, Freud18}.
To see the computational complexity,
we use ``tic" and ``toc" commands in MATLAB
and 
show the results in Fig.~\ref{Fig:plt_times}
as functions of key parameters
when $(\mu, N_{\rm run}) = (0.5, 10)$ and SNR $= 10$ dB.
Since the complexity
is independent of the number of active transmit antennas, $K$,
the computing time is almost invariant with respect to $K$ in 
Fig.~\ref{Fig:plt_times} (a),
while we can observe that
the computing time grows 
quadratically with the number of receive antennas, $N$,
and linearly with 
the number of transmit antennas, $L$, in 
Fig.~\ref{Fig:plt_times} (b) and (c), respectively.
From Fig.~\ref{Fig:plt_times},
we can also confirm that
due to the exhaustive search for
the ML detection in \eqref{EQ:ml_ga}, in general,
the complexity of the ML detection is much higher than
that of the CAVI algorithm (except that $K$ is sufficiently small,
i.e., $K \le 2$ as mentioned earlier).
Thus, the CAVI algorithm can be seen as a low-complexity
approach to the ML detection in \eqref{EQ:ml_ga} and \cite{Lin15,
Freud18} with negligible performance degradation (as shown
in Figs.~\ref{Fig:plt3} and~\ref{Fig:plt4}).

\section{Concluding Remarks}	\label{S:Conc}

In this paper, we considered a simple transmit coding
scheme for SM, namely RCSM, and studied
the index detection problem using the Gaussian approximation,
which led to an approximate ML formulation for the index detection.
Based on the derived approximate closed-form expression for the probability
of index error, we showed that the solution to
the approximate ML formulation can still have a full 
coding gain.
Since an exhaustive search to find the solution to 
the ML problem required a prohibitively high
complexity, which grows exponentially
with the number of active transmit antennas, 
we applied a variational inference approach,
which is a well-known machine learning approach, and derived 
an iterative algorithm. The resulting iterative
algorithm has a low complexity thanks to the Gaussian approximation.
In particular, it was shown that the complexity 
of the proposed algorithm
is independent of the number of active transmit antennas. 
From simulation results and 
the approximate closed-form expression,
we also saw that the proposed iterative
algorithm can achieve a near ML performance,
which demonstrates the benefit of machine learning algorithms to
the signal detection in SM.


\appendices

\section{Proof of Lemma~\ref{L:Ineq2}}	\label{A:Ineq2}
Consider the expectation of $\bV(x)$ over $x_t$,
which is given by
\begin{align}
& \uE_t [\bV(\bx)] = 
\uE_t \left[\left( \bh_t \bh_t^\rH x_t +
\bA_t\right)^{-1} \right] \cr
& \quad = \bA_t^{-1} 
- \uE_t \left[
\frac{x_t}{1 + x_t \bh_t^\rH \bA_t^{-1} \bh_t}
\right] \bA_t^{-1} \bh_t \bh_t^\rH \bA_t^{-1},  
\end{align}
where 
$\bA_t = 
\sum_{p \ne t} \bh_p \bh_p^\rH x_p + \gamma^{-1} \bI$ and
the second equality is due to the matrix inversion lemma
\cite{Harv97}.
Since $\Pr(x_t = 1) = q_t$ and 
$\Pr(x_t = 0) = 1 - q_t$, 
it can be shown that
\begin{align}
\uE_t \left[
\frac{x_t}{1 + x_t \bh_t^\rH \bA_t^{-1} \bh_t}
\right]
= \frac{q_t}{1 + \bh_t^\rH \bA_t^{-1} \bh_t} 
\le \frac{q_t}{1 + q_t \bh_t^\rH \bA_t^{-1} \bh_t},
	\label{EQ:Eqq}
\end{align}
which results in $\uE_t [\bV(\bx)] \succeq
\left( \bh_t \bh_t^\rH q_t + \bA_t
\right)^{-1}$.
Therefore, 
if the expectation is carried out over $t \in \{1, \ldots, l-1,
l+1, \ldots, L\}$,
we can have the following result:
$$
\uE_{-l} [\bV(\bx)] 
\succeq  
\left( \bh_l \bh_l^\rH x_l
+ \sum_{t \ne l}
\bh_t \bh_t^\rH q_t + \gamma^{-1} \bI \right)^{-1},
$$
which proves the first equation in \eqref{EQ:L1}.

According to the matrix determinant lemma \cite{Harv97},
we have
\begin{align}
\uE_t [\phi(\bx)] & = 
\uE_t\left[\ln \det 
\left( \sum_{p=1}^L \bh_p \bh_p^\rH x_p + \gamma^{-1} \bI
\right) \right] \cr
& = \uE_t[ \ln (1+ x_t \bh_t^\rH \bA_t^{-1} \bh_t) ] + \ln \det(\bA_t) 
 \cr
& \le \ln (1+ q_t \bh_t^\rH \bA_t^{-1} \bh_t) + \ln \det(\bA_t)  \cr
& = \ln \det
\left( \bh_t \bh_t^\rH q_t +\sum_{p \ne t}^L \bh_p \bh_p^\rH x_p 
+ \gamma^{-1} \bI \right) ,
	\label{EQ:lnd}
\end{align}
where the inequality is due to Jensen's inequality.

In order to see the inequalities in 
\eqref{EQ:L1} are tight as $q_t \to 0$ or $1$ for
$t \in \{1, \ldots, l-1,l+1, \ldots, L\}$,
consider \eqref{EQ:Eqq}.
It can be shown that
$$
\frac{q_t}{1 + q_t \bh_t^\rH \bA_t^{-1} \bh_t}-
\frac{q_t}{1 + \bh_t^\rH \bA_t^{-1} \bh_t} 
= \frac{q_t (1-q_t) d_t}{(1+ q_t d_t)(1+d_t)},
$$
where $d_t = \bh_t^\rH \bA_t^{-1} \bh_t$.
Thus, as $q_t$ approaches $0$ or $1$, the 
inequality in \eqref{EQ:Eqq} becomes tight. As a result,
the first inequality in \eqref{EQ:L1} becomes tight.
Similarly, we can also show that 
the inequality in \eqref{EQ:lnd} is tight 
if $q_t$ approaches $0$ or $1$. Thus, 
the second inequality in \eqref{EQ:L1} becomes tight for
$q_t \to 0$ or $1$. 

\section{Proof of Lemma~\ref{L:PEP}}	\label{A:PEP}

Using the Chernoff bound \cite{CoverBook}, 
it can be shown that
\begin{align}
& \Pr\left(\sum_{m=1}^M \by_m^\rH \bDelta \by_m  > d\,|\, 
\bh_1, \bh_2, \bD \right) \cr
& \le
e^{-\lambda d}
 \uE \left[e^{\lambda \sum_{m=1}^M \by_m^\rH \bDelta \by_m}\,|\, 
\bh_1, \bh_2, \bD \right] \cr
& = e^{-\lambda d} \prod_{m=1}^M
 \uE \left[e^{\lambda \by_m^\rH \bDelta \by_m}\,|\, 
\bh_1, \bh_2, \bD \right],
\end{align}
where $\lambda > 0$.
Let 
$\hat \by_m = s_{m,1}^* \by_m$. 
From \eqref{EQ:ym1}, under the assumption of {\bf A1}
and the Gaussian approximation for the interference-plus-noise
in \eqref{EQ:ym1},
we can show that
$\hat \by_m = \bh_1 + s_{m,1}^* \bee_m$,
where $s_{m,1}^* \bee_m \sim \cC \cN(\b0, \bD)$.
Then, since 
$\by_m^\rH \bDelta \by_m = \hat \by_m^\rH \bDelta \hat \by_m$,
after some manipulations, we have
\begin{align}
& \uE \left[e^{\lambda \by_m^\rH \bDelta \by_m}\,|\, 
\bh_1, \bh_2, \bD \right] \cr
& =
\frac{1}{\pi^N{\rm det}(\bD)} \int  
e^{\lambda \hat \by_m^\rH \bDelta \hat \by_m}
e^{-(\hat \by_m - \bh_1)^\rH \bD^{-1} (\hat \by_m - \bh_1)} d \by_m \cr
& = \frac{{\rm det}(\bW^{-1})}{{\rm det} (\bD)}
\exp \left(
-\bh_1^\rH (\bD^{-1} - \bD^{-1} \bW^{-1} \bD^{-1} ) \bh_1
\right),
	\label{EQ:Ee}
\end{align}
where $\bW = \bD^{-1} - \lambda \bDelta$.
Using the matrix inversion lemma,
we can show that
\begin{align}
\bDelta = \bD^{-1} (\tau_2 \bh_2 \bh_2^\rH - 
\tau_1 \bh_1 \bh_1^\rH ) \bD^{-1},
	\label{EQ:Dtt}
\end{align}
where $\tau_l = \frac{1}{1+ \alpha_l}$.
According to \eqref{EQ:Ee},
$\kappa (\lambda)$ is given by
\begin{align}
\kappa (\lambda)
= \frac{{\rm det}(\bW^{-1})}{{\rm det}(\bD)} 
= \frac{1}{{\rm det}(\bI  - \lambda \bD \bDelta)}.
	\label{EQ:k2}
\end{align}
Using the matrix determinant lemma \cite{Harv97},
we have
\begin{align}
{\rm det}(\bI  - \lambda \bD \bDelta)
& = {\rm det}(\bB) (1 + \lambda 
\tau_1 \bh_1^\rH \bD^{-1} \bB^{-1} \bh_1),
	\label{EQ:dB}
\end{align}
where $\bB = \bI - \lambda \tau_2 \bh_2 \bh_2^\rH$.
Using the matrix determinant lemma,
it can be shown that
\be
{\rm det}(\bB) 
= 1 - \lambda \tau_2 \alpha_2 = \frac{\lambda}{
(1+ \alpha_2) \theta_2 (\lambda)}.
	\label{EQ:k1t}
\ee
Applying the matrix inversion lemma, it can be shown that
$$
\bB^{-1} = \bI + \frac{ \lambda \tau_2 }{1 - \lambda \tau_2 \alpha_2}
\bh_2 \bh^\rH_2 \bD^{-1}.
$$
From this, we have
\begin{align}
\lambda \tau_1 \bh_1^\rH \bD^{-1} \bB^{-1} \bh_1
= \lambda \tau_1 \left(
\alpha_1 + \theta_2 (\lambda) |\beta|^2.
\right)
	\label{EQ:k2t}
\end{align}
Substituting \eqref{EQ:k1t} and \eqref{EQ:k2t}
into \eqref{EQ:dB},
it can be shown that
\begin{align}
{\rm det}(\bI  - \lambda \bD \bDelta)
= \frac{\lambda}{(\alpha_2 + 1) \theta_2 (\lambda)}
\left(
1 + \frac{\lambda (\alpha_1 +\theta_2 (\lambda) |\beta|^2)}{\alpha_1 + 1}
\right).
	\label{EQ:dd}
\end{align}
Substituting \eqref{EQ:dd} into \eqref{EQ:k2},
we can show the first equation in \eqref{EQ:alal}.

Since
$\bW = \bD^{-1} (\bD - \lambda \bD  \bDelta \bD) \bD^{-1}$,
it can be shown that
\begin{align}
& \bh_1^\rH (\bD^{-1} - \bD^{-1} \bW^{-1} \bD^{-1} ) \bh_1 \cr
& = \alpha_1 - \bh_1^\rH (\bD - \lambda \bD \bDelta \bD)^{-1} \bh_1.
\label{EQ:hx}
\end{align}
Substituting \eqref{EQ:Dtt} into \eqref{EQ:hx}
and using the matrix inversion lemma,
we have
\begin{align}
\bh_1^\rH (\bD^{-1} - \bD^{-1} \bW^{-1} \bD^{-1} ) \bh_1 
= \alpha_1 - \frac{c_1}{1 + \lambda \tau_2  c_1},
	\label{EQ:psinr}
\end{align}
where
\begin{align}
c_1 & = \bh_1^\rH (\bD - \lambda \tau_2 \bh_2 \bh_2)^{-1} \bh_1 \cr
& = \alpha_1 + 
\frac{\lambda \tau_2}{1 - \lambda \tau_2 \alpha_2}
|\bh_1^\rH \bD^{-1} \bh_2|^2 
= \alpha_1 + \theta_2 (\lambda) |\beta |^2,
	\label{EQ:c_1}
\end{align}
since $\theta_2 (\lambda) = 
\frac{\lambda \tau_2}{1 - \lambda \tau_2 \alpha_2}$.
Substituting \eqref{EQ:c_1} into \eqref{EQ:psinr},
we have
\begin{align}
\gamma_{\rm pep} (\lambda)
& = \bh_1^\rH (\bD^{-1} - \bD^{-1} \bW^{-1} \bD^{-1} ) \bh_1  \cr
& = 
\alpha_1 - 
\frac{(\alpha_1 +1) (\alpha_1 + \theta_2(\lambda) |\beta|^2)}{
\alpha_1 +1+ \lambda (\alpha_1 + \theta_2(\lambda) |\beta|^2)},
\end{align}
which is the second equation in \eqref{EQ:alal}.

\section{Proof of Lemma~\ref{L:Asym}}	\label{A:Asym}
In \eqref{EQ:alal}, for large $N$ and $L$
with fixed $\eta = Q/N$, 
$\alpha_l$ 
and $|\beta|^2$ 
approach $\bar \alpha = \uE[\alpha_l]$ and
$\uE[|\beta|^2]$, respectively.
Thus, the asymptotic PEP-SNR can be obtained
by replacing 
$\alpha_l$ and $|\beta|^2$ 
with $\bar \alpha$,
which is given in \eqref{EQ:balpha} and
$\uE[|\beta|^2]$, respectively.

For asymptotic $|\beta|^2$, from \cite{TulinoBook},
we have
\begin{align}
|\beta|^2 & \to  \uE[|\beta|^2] \cr
& =  \uE[\bh_1^\rH \bD^{-1} \bh_2 \bh_2^\rH \bD^{-1} \bh_1] 
=  \uE[\bh_1^\rH \bD^{-1} \uE[\bh_2 \bh_2^\rH] \bD^{-1} \bh_1] \cr
& = \frac{1}{N} \uE[\bh_1^\rH \bD^{-2} \bh_1] 
= \frac{1}{N} \left( \frac{ \uE[{\rm tr}(\bD^{-2})]}{N} \right).
\end{align}
It can be shown that
\begin{align*}
\bD^2 
& = (\gamma^{-1} \bI +  \bar \bH \bar \bH^\rH)^2 
= \gamma^{-2} \bI + 2 \gamma^{-1}  \bar \bH \bar \bH^\rH
+ \bar \bH \bar \bH^\rH\bar \bH \bar \bH^\rH \cr
& \approx
\gamma^{-2} \bI + 2 \gamma^{-1}  \bar \bH \bar \bH^\rH
+ \bar \bH \bar \bH^\rH \cr
& = (2 \gamma^{-1} + 1) 
\left( \frac{1}{\gamma^2 + 2 \gamma} \bI + \bar \bH \bar \bH^\rH \right),
\end{align*}
where
the approximation is due to $\bH^\rH \bH \approx \bI$
for a large $N$ and fixed $\eta = \frac{Q}{N}$.
Then, it can be shown that
\begin{align}
\frac{ \uE[{\rm tr}(\bD^{-2})]}{N} 
& \approx
\frac{ \uE\left[
\left(
\frac{1}{\gamma^2 + 2 \gamma} \bI + \bar \bH \bar \bH^\rH 
\right)^{-1}
\right] }
{2 \gamma^{-1} + 1} \cr
& = 
\frac{
\gamma_{\rm mmse} \left( \gamma (2 + \gamma), \eta \right)}
{2 \gamma^{-1} + 1} = \bar \omega.
\end{align}
From this, we have $\uE[|\beta|^2] \approx
\frac{\bar \omega}{N}$,
which is substituted into the PEP-SNR expression
in \eqref{EQ:alal}.
Then, we can obtain \eqref{EQ:ub_SNR}.

\bibliographystyle{ieeetr}
\bibliography{mimo}

\end{document}